\documentclass[lettersize,hidelinks,journal]{IEEEtran}

\usepackage{cite}
\usepackage{amsmath,amssymb,amsfonts}
\usepackage{graphicx}
\usepackage{textcomp}
\usepackage[dvipsnames]{xcolor}
\usepackage[acronym,shortcuts]{glossaries}
\usepackage{bm}
\usepackage{caption}
\usepackage{subcaption}
\usepackage{relsize}
\usepackage{soul}
\usepackage{algorithm, algpseudocode}
\usepackage{algcompatible}
\usepackage{outlines}
\usepackage{orcidlink}
\usepackage{lipsum,comment}
\usepackage{mathtools}
\usepackage{tabularx,flushend}
\newtheorem{remark}{\hspace{-2.5ex}\bf Remark}

\def\BibTeX{{\rm B\kern-.05em{\sc i\kern-.025em b}\kern-.08em
T\kern-.1667em\lower.7ex\hbox{E}\kern-.125emX}}

\newcommand{\vect}[1]{\mathrm{vec}\!\left({#1}\right)}

\newcommand{\trans}[0]{^{\mathsf{T}}}

\newcommand{\herm}[0]{^{\mathsf{H}}}

\newcommand{\Real}[1]{\Re\{{#1}\}}
\newcommand{\Imag}[1]{\Im\{{#1}\}}

\newacronym{RPE}{RPE}{radar parameter estimation}
\newacronym{OTFS}{OTFS}{orthogonal time frequency space}
\newacronym{AFDM}{AFDM}{affine frequency division multiplexing}
\newacronym{MIMO}{MIMO}{multiple-input multiple-output}
\newacronym{SISO}{SISO}{single-input single-output}
\newacronym{ISAC}{ISAC}{integrated sensing and communications}
\newacronym{3D}{3D}{three-dimensional}
\newacronym{2D}{2D}{two-dimensional}
\newacronym{1D}{1D}{one-dimensional}
\newacronym{RX}{RX}{receiver}
\newacronym{TX}{TX}{transmitter}
\newacronym{BF}{BF}{beamforming}
\newacronym{mmWave}{mmWave}{millimeter-wave}
\newacronym{SotA}{SotA}{state-of-the-art}
\newacronym{ULA}{ULA}{uniform linear array}
\newacronym{QAM}{QAM}{quadrature amplitude modulation}
\newacronym{ISFFT}{ISFFT}{inverse symplectic finite Fourier transform}
\newacronym{SFFT}{SFFT}{symplectic finite Fourier transform}
\newacronym{AWGN}{AWGN}{additive white Gaussian noise}
\newacronym{OFDM}{OFDM}{orthogonal frequency division multiplexing}
\newacronym{OCDM}{OCDM}{orthogonal chirp division multiplexing}
\newacronym{BS}{BS}{base station}
\newacronym{UE}{UE}{user equipment}
\newacronym{DFT}{DFT}{discrete Fourier transform}
\newacronym{IDFT}{IDFT}{inverse discrete Fourier transform}
\newacronym{TD}{TD}{time-domain}
\newacronym{wlg}{wlg}{without loss of generality}
\newacronym{CP}{CP}{cyclic prefix}
\newacronym{DAFT}{DAFT}{discrete affine Fourier transform}
\newacronym{IDAFT}{IDAFT}{inverse discrete affine Fourier transform}
\newacronym{CPP}{CPP}{\textit{chirp-periodic} prefix}
\newacronym{IDZT}{IDZT}{inverse discrete Zak transform}
\newacronym{DZT}{DZT}{discrete Zak transform}
\newacronym{ICI}{ICI}{inter-carrier interference}
\newacronym{BER}{BER}{bit error rate}
\newacronym{DoF}{DoF}{degrees-of-freedom}
\newacronym{FD}{FD}{full-duplex}
\newacronym{SIMO}{SIMO}{single-input multiple-output}
\newacronym{MISO}{MISO}{multiple-input single-output}
\newacronym{AoD}{AoD}{angle-of-departure}
\newacronym{AoA}{AoA}{angle-of-arrival}
\newacronym{RF}{RF}{radio frequency}
\newacronym{SIM}{SIM}{stacked intelligent metasurfaces}
\newacronym{FPGA}{FPGA}{field programmable gate array}
\newacronym{UPA}{UPA}{uniform planar array}
\newacronym{CC}{CC}{communication-centric}
\newacronym{I/O}{I/O}{input-output}
\newacronym{iid}{i.i.d.}{independent and identically distributed}
\newacronym{IoT}{IoT}{internet of things}
\newacronym{V2X}{V2X}{vehicle-to-everything}
\newacronym{NTN}{NTN}{non-terrestrial network}
\newacronym{LEO}{LEO}{low-earth orbit}
\newacronym{THz}{THz}{terahertz}
\newacronym{EM}{EM}{electromagnetic}
\newacronym{RIS}{RIS}{reconfigurable intelligent surface}
\newacronym{DoA}{DoA}{direction-of-arrival}
\newacronym{DD}{DD}{doubly-dispersive}
\newacronym{ODDM}{ODDM}{orthogonal delay-Doppler division multiplexing}
\newacronym{LoS}{LoS}{line-of-sight}
\newacronym{NLoS}{NLoS}{non-line-of-sight}
\newacronym{6G}{6G}{sixth generation}
\newacronym{MPDD}{MPDD}{metasurfaces-parametrized DD}
\newacronym{GaBP}{GaBP}{Gaussian Belief Propagation}
\newacronym{MSE}{MSE}{mean-squared-error}
\newacronym{sIC}{soft IC}{soft interference cancellation}
\newacronym{soft RG}{soft RG}{soft replica generation}
\newacronym{BG}{BG}{belief generation}
\newacronym{SGA}{SGA}{scalar Gaussian approximation}
\newacronym{CLT}{CLT}{central limit theorem}
\newacronym{PDF}{PDF}{probability density function}
\newacronym{QPSK}{QPSK}{quadrature phase-shift keying}
\newacronym{LMMSE}{LMMSE}{linear minimum mean square error}
\newacronym{SNR}{SNR}{signal-to-noise ratio}
\newacronym{ZF}{ZF}{zero-forcing}
\newacronym{PDA}{PDA}{probabilistic data association}

\begin{document}

\title{Doubly-Dispersive MIMO Channel{s \\with} Stacked Intelligent Metasurfaces{:\\ Modeling, Parametrization, and Receiver Design}}

\author{Kuranage Roche Rayan Ranasinghe\textsuperscript{\orcidlink{0000-0002-6834-8877}},~\IEEEmembership{Graduate Student Member,~IEEE,}\\
Iv{\'a}n Alexander Morales Sandoval\textsuperscript{\orcidlink{0000-0002-8601-5451}},~\IEEEmembership{Graduate Student Member,~IEEE,}
Hyeon Seok Rou\textsuperscript{\orcidlink{0000-0003-3483-7629}},~\IEEEmembership{Member,~IEEE,}\\
Giuseppe Thadeu Freitas de Abreu\textsuperscript{\orcidlink{0000-0002-5018-8174}},~\IEEEmembership{Senior Member,~IEEE,}\\ and
George C. Alexandropoulos\textsuperscript{\orcidlink{0000-0002-6587-1371}},~\IEEEmembership{Senior Member,~IEEE}
\thanks{K. R. R. Ranasinghe, I. A. M. Sandoval, H. S. Rou and G. T. F. de Abreu are with the School of Computer Science and Engineering, Constructor University (previously Jacobs University Bremen), Campus Ring 1, 28759 Bremen, Germany (emails: \{kranasinghe, imorales, hrou, gabreu\}@constructor.university).} 
\thanks{G. C. Alexandropoulos is with the Department of Informatics and Telecommunications, National and Kapodistrian University of Athens, 15784 Athens, Greece and with the Department of Electrical and Computer Engineering, University of Illinois Chicago, Chicago, IL 60601, USA (e-mail: alexandg@di.uoa.gr).
}\vspace{-3ex}}



\maketitle

\begin{abstract}
Introduced with the advent of statistical wireless channel models for high mobility communications and having a profound role in \ac{CC} \ac{ISAC}, the \ac{DD} channel structure has long been heralded as a useful tool enabling the capture of the most important fading effects undergone by an arbitrary time-domain  transmit signal propagating through some medium. 
However, the incorporation of this model into \ac{MIMO} system setups, relying on the recent paradigm-shifting transceiver architecture based on \ac{SIM}, in an environment with  \acp{RIS} remains an open problem due to the many intricate details that have to be accounted for. 
In this paper, we fill this gap by introducing a novel \ac{DD} \ac{MIMO} channel model that incorporates an arbitrary number of \acp{RIS} in the ambient, as well as \acp{SIM} equipping both the transmitter and receiver. 
%
%
We then discuss how the proposed \ac{MPDD} channel model can be seamlessly applied to waveforms that are known to perform well in \ac{DD} environments, namely, \ac{OFDM}, \ac{OTFS}, and \ac{AFDM}, with each having their own inherent advantages and disadvantages. 
An illustrative application of the programmable functionality of the proposed model is finally presented to showcase its potential for boosting the performance of the aforementioned waveforms. 
Our numerical results indicate that the design of waveforms suitable to mitigating the effects of \ac{DD} channels is significantly impacted by the emerging \ac{SIM} technology.
%
\end{abstract}


\glsresetall

\vspace{-2ex}
\section{Introduction}

\IEEEPARstart{N}{ext} generation wireless communications systems are expected to bring about a plethora of functionalities and support for applications that insofar have not been feasible, such as \ac{V2X} and aerial communications \cite{ChenCSM2017}, \ac{IoT} networks \cite{NguyenIoTJ2022}, and \acp{NTN} such as \ac{LEO} satellite networks \cite{ShiNetwork2024}, all of which require robustness against high mobility scenarios
\cite{LiCOMMST2022}. 

Traditionally, high mobility communication scenarios are known to pose significant challenges \cite{WangTWC2006} due to the time-frequency selectivity present in time-varying multipath conditions, giving rise to the necessity of using \ac{DD} channel models \cite{LiuTIT2004, Bliss_Govindasamy_2013, Rou_SPM_2024}, whose properties have been recently leveraged for various purposes \cite{SurabhiTWC2019, BomfinTWC2021,PfadlerTWC2024, LiangTWC2024, HaifTVT2024,ZhangJSAC2024}. 

A notable example of the latter is the exploitation of \ac{DD} channel models for the intrinsic capturing of radar-like parameters, enabling \ac{RPE} over communication waveforms, giving rise to \ac{CC} \ac{ISAC}\footnote{We distinguish \ac{CC} \ac{ISAC} from other coexistence approaches, which either try to communicate over radar waveforms or involve the joint design of both communication and radar subsystems \cite{Liu_JSAC_2022, CSL2023, GonzalezProcIEEE2024, FD_MIMO_ISAC_2024}.} \cite{HyeonTWC2024, GaudioTWC2020, Mohammed_BITS_2022, Gupta_OJCS_2024, Ranasinghe_ICASSP_2024, KuranageTWC2024}.
A particularly motivating aspect of \ac{CC}-\ac{ISAC} is that the combination of \ac{RPE} methods with  techniques commonly used in the processing of communication signals, such as the de-chirping of \ac{AFDM} signals \cite{Bemani_WCL_2024}, bilinear inference \cite{Parker_TSP_2014, IimoriTWC2021, TakahashiTWC2023} and blind covariance-based detection \cite{YangCommL2011,BaoACCESS2019}, enables implementation of \ac{ISAC} in mono-, bi- and multi-static fashions \cite{RanasingheWCNC2024}.

However, \ac{DD} channels require careful waveform design, with some main contenders being \ac{OFDM} \cite{Gupta_OJCS_2024}, \ac{OTFS} \cite{GaudioTWC2020, Mohammed_BITS_2022, Ranasinghe_ICASSP_2024, Gupta_OJCS_2024} and \ac{AFDM} \cite{Ni_ISWCS_2022, Bemani_TWC_2023, RouAsilimoar2024}.
For example, the widely adopted \ac{OFDM} waveform, despite having features that are useful for \ac{RPE} \cite{LiyanaarachchiTWC2024}, suffers from high inter-carrier interference that reduces robustness to high Doppler shifts present in \ac{DD} environments, leading to severe performance degradation in high mobility scenarios \cite{Gaudio_TWC_2022}. 
In turn \ac{OTFS}, which is a \ac{2D} modulation scheme that directly embeds information on the delay-Doppler domain, offers an alternative to \ac{OFDM} but requires a significantly higher implementation complexity than the latter, and was found not to achieve optimal diversity order in \ac{DD} channels \cite{SurabhiTWC2019}.
Finally, \ac{AFDM}, which has been recently proposed \cite{Ni_ISWCS_2022, Bemani_TWC_2023, RouAsilimoar2024} aiming to address the aforementioned weakness of \ac{OTFS}, has been shown to achieve optimal diversity order in \ac{DD} channels but may also face implementation challenges related to the generation of chirp signals at high frequencies \cite{Srivastava_ISES2018}.

%
%
Fortunately, it has been recently shown that, from a mathematical standpoint, the \ac{DD} channel model for \ac{OFDM}, \ac{OTFS} and \ac{AFDM} -- as well as several other related waveforms including \ac{OCDM} \cite{OuyangTCom2016} and \ac{ODDM} \cite{TongTCom2024} -- have a similar structure, which enables the design of systems for these waveforms in a unified manner \cite{Rou_SPM_2024}.
The unified model of \cite{Rou_SPM_2024} is, however, limited to the \ac{SISO} case, while the prominent role of the \ac{MIMO} technology in the physical-layer of current and future wireless networks has been well established, including recent variations such as extremely large \ac{MIMO} \cite{XLMIMO_tutorial, NF_beam_tracking} and technologies designed around reconfigurable metasurfaces \cite{Tsinghua_RIS_Tutorial, BAL2024}, which are considered for the upcoming \ac{6G} systems.

All the above motivates us to consider a \ac{DD}-\ac{MIMO} channel model incorporating multi-functional reconfigurable metasurfaces, including both \ac{RIS} \cite{LiuCommST2021, Tsinghua_RIS_Tutorial, BAL2024} and the recently proposed \ac{SIM} \cite{PG13,ZKW+18,HLC+19}, to enable the design of advanced \ac{6G} systems capable of supporting high mobility.
{
Indeed, although a growing body of work on \acp{SIM} is building, with topics such as channel estimation \cite{YaoWCL2024} and \ac{ISAC} \cite{NiuWCL2024,LiTVT2025} being well covered, contributions so far are typically limited to sub-6GHz channel model, falling short of incorporating mobility scenarios and \ac{DD} channels. 
}

{We therefore introduce in this article} a novel \ac{MPDD} \ac{MIMO} channel model incorporating \acp{SIM} at both the \ac{TX} and \ac{RX} as well as multiple \acp{RIS} in the environment. 
Taking into account both the conventional hybrid analog and digital \ac{BF} framework in \cite{SrivaTWC2022} and the \ac{EM}-compliant \ac{BF} framework for \acp{RIS} and \acp{SIM} \cite{AnWC2024}, we derive the end-to-end \ac{I/O} relationship for arbitrary \ac{TD} transmit signals passing through the proposed \ac{DD} channel model. 
We also extend the \ac{TD} relationship to encompass the end-to-end \ac{I/O} relationships with the \ac{OFDM}, \ac{OTFS}, and \ac{AFDM} waveforms and showcase the differences in their effective channels. 
Finally, we present an optimization example of the proposed programmable \ac{MPDD} \ac{MIMO} channel model. 
{Our contributions} are summarized as follows:
\begin{itemize}
\item {A} novel point-to-point MPDD channel model {extending that \cite{Rou_SPM_2024} to \ac{MIMO} scenarios} including \ac{TX} and \ac{RX} \acp{SIM} {and} \acp{RIS}, which is suitable for high-mobility scenarios and \ac{CC}-\ac{ISAC} {is described}.    
\item Capitalizing on the {above}, novel expressions for the TD received signal as well as the effective channel matrices with \ac{OFDM}, \ac{OTFS}, and \ac{AFDM} waveforms are presented{, which offer insights into the features of reconfigurable electromagnetic technologies applied to \ac{DD} systems, in addition to enabling} the formulation of various \ac{ISAC} objectives.
\item As an application example, we describe and solve an optimization problem whereby \acp{SIM} located at the transmitter and receiver are programmed to enhance receive signal power and therefore boost the detection performance of \ac{OFDM}, \ac{OTFS}, and \ac{AFDM} waveforms in a \ac{MPDD}-\ac{MIMO} channel.
To that end, besides the closed-form expressions for the gradient of the objective function of such a problem, we also design a purpose-built detector based on the \ac{GaBP} technique.
Simulation results demonstrate significant gains due to the optimized \acp{SIM} over a conventional \ac{DD} model without the parametrized metasurfaces, resulting in a robust reduction in the performance gap between the classic \ac{OFDM} scheme compared to the more modern \ac{OTFS} and \ac{AFDM} waveforms.
\end{itemize}

\textit{Notation:} All scalars are represented by upper or lowercase letters, while column vectors and matrices are denoted by bold lowercase and uppercase letters, respectively.
The diagonal matrix constructed from vector $\mathbf{a}$ is denoted by diag($\mathbf{a}$), while $\mathbf{A}\trans$, $\mathbf{A}\herm$, $\mathbf{A}^{1/2}$, and $[\mathbf{A}]_{i,j}$ denote the transpose, Hermitian, square root and the $(i,j)$-th element of a matrix $\mathbf{A}$, respectively.
The convolution and Kronecker product are respectively denoted by $*$ and $\otimes$, while $\mathbf{I}_N$ and $\mathbf{F}_N$ represent the $N\times N$ identity and the normalized $N$-point \ac{DFT} matrices, respectively.
The sinc function is expressed as $\text{sinc}(a) \triangleq \frac{\sin(\pi a)}{\pi a}$, and $\jmath\triangleq\sqrt{-1}$ denotes the elementary complex number.
%

\vspace{-2ex}
\section{Preliminaries}

\subsection{Antenna Array Response}

Let $\phi\in [0,\pi]$ denote the arbitrary \ac{AoA} or \ac{AoD} of a channel propagation path to (or from) a \ac{ULA} with $A$ antenna elements. Then, the array response vector $\mathbf{a} (\phi) \in \mathbb{C}^{A \times 1}$ is defined as
\begin{equation}
\label{eq:ULA_response_vector_transmit}
\mathbf{a} (\phi) \triangleq \tfrac{1}{\sqrt{A}} \Big[ 1, e^{-\jmath\frac{2\pi}{\lambda}d \sin(\phi)}, \dots, e^{-\jmath\frac{2\pi}{\lambda}(A-1)d \sin(\phi)} \Big]\trans,
\end{equation}
where $\lambda$ indicates the wavelength and $d$ is the antenna spacing, which is usually set as $d=\lambda/2$ \cite{SrivaTWC2022}.

Similarly, for a \ac{UPA} with $B \triangleq B_x B_z$ elements\footnote{Without loss of generality, the \ac{UPA} is aligned parallel to the $y$ direction with elements occupying space in the $x$ and $z$ dimensions. The generalization to arbitrary axes is trivial; some other orientations are discussed in \cite{AnJSAC2024}.
}, the response vector corresponding to a path impinging onto (or outgoing from) the array at the elevation and azimuth angles $\theta\in [0,\pi]$ and $\phi \in [-\frac{\pi}{2},\frac{\pi}{2}]$, is given by \cite{AnJSAC2023}
\begin{equation}
\label{eq:SIM_transmit_UPA_response vector}
\mathbf{b}(\phi,\theta) \triangleq \tfrac{1}{\sqrt{B_x B_z}}  \mathbf{b}_x(\phi,\theta) \otimes \mathbf{b}_z(\theta) \in \mathbb{C}^{B \times 1},
\end{equation}
where the $x$- and $z$-axis steering vectors $\mathbf{b}_x(\phi,\theta) \in \mathbb{C}^{B_x \times 1}$ and $\mathbf{b}_z(\theta) \in \mathbb{C}^{B_z \times 1}$ are respectively defined as
\begin{subequations}
\vspace{-1ex}
\begin{eqnarray}
\label{eq:steering_x_axis} 
\mathbf{b}_x(\phi,\theta) \triangleq \Big[1, e^{-\jmath \frac{2\pi d_x}{\lambda} \sin(\phi) \sin(\theta)}, \dots,&& \\[-1ex]
&&\hspace{-10ex} e^{-\jmath \frac{2\pi d_x}{\lambda} (B_x - 1)  \sin(\phi) \sin(\theta)} \Big]\trans,
\nonumber
\end{eqnarray}
\vspace{-4ex}

\noindent and 
\begin{equation}
\label{eq:steering_z_axis} 
\!\!\mathbf{b}_z(\theta) \triangleq \!\Big[1, e^{-\jmath \frac{2\pi d_z}{\lambda} \cos(\theta)}, \dots, e^{-\jmath \frac{2\pi d_z}{\lambda} (B_z - 1) \cos(\theta)} \Big]\trans,
\end{equation}
\end{subequations}
with $d_x$ and $d_z$ being the element spacing in the \ac{UPA}'s $x$- and $z$-axis directions, respectively, which are usually set as $d_x = d_z = \lambda/2$.

\vspace{-2ex}
\subsection{SIM Modeling}
\label{sec:SIM_Modeling}

Consider a \ac{SIM} with $Q$ layers of transmissive metasurfaces placed in parallel at very close distances, where each metasurface consists of $M \triangleq M_x M_z$ response-tunable meta-atoms\footnote{For simplicity, we consider the homogeneous case. The extension to a heterogeneous variation, with metasurfaces containing different numbers of atoms, is trivial but notationally laborious with no fundamental insight gained.} with $M_x$ and $M_z$ denoting the number of meta-atoms in the $x$- and $z$-axis on each layer, respectively. 
\newpage

We define the following $M \times M$ matrix including the effective tunable phase shifts of all $M$ meta-atoms embedded in each $q$-th metasurface layer, with $q \in \mathcal{Q}\triangleq \{ 1, \dots, Q \}$, of the \ac{SIM} as
%
\begin{equation}
\bm{\Psi}_q  \triangleq \text{diag}\bigg( \!\Big[ e^{\jmath\zeta^{q}_{1}}, \dots, e^{\jmath\zeta^{q}_{M}} \Big]\! \bigg),
\label{eq:diagona_shift_matrix_per_layer}
\end{equation}
where $\zeta^{q}_{m} \in [0,2\pi)$ $\forall q \in \mathcal{Q}$ and $\forall m \in \mathcal{M} \triangleq \{ 1, \dots, M \}$ represents the transmissive phase response of the $m$-th meta-atom lying on the $q$-th metasurface layer. 

The transmission matrix between each $(q-1)$-th and the $q$-th layer of the \ac{SIM} $\forall q \in \mathcal{Q} \backslash \{ 1\}$ is denoted by $\bm{\Gamma}_q \in \mathbb{C}^{M \times M}$. 
According to the Rayleigh-Sommerfeld diffraction theory, each $(m,m')$-th element (with $m,m' \in \mathcal{M}$) of $\bm{\Gamma}_q$ represents the diffraction coefficient between the $m'$-th meta-atom on the $(q-1)$-th metasurface and the $m$-th meta-atom on the $q$-th metasurface, and is given by \cite{AnJSAC2023}
\begin{equation}
\label{eq:diffraction_coeff}
\gamma^{q}_{m,m'} \triangleq \frac{\rho_t \cos\big(\epsilon^{q}_{m,m'}\big)}{\mathrm{d}^{q}_{m,m'}}  \bigg( \frac{1}{2\pi  \mathrm{d}^{q}_{m,m'}} - \frac{\jmath}{\lambda} \bigg)  e^{\jmath2\pi \frac{\mathrm{d}^{q}_{m,m'}}{\lambda}},
\end{equation}
where $\rho_t$ denotes the square measure occupied by each meta-atom in the \ac{SIM}, $\epsilon^{q}_{m,m'}$ is the angle between the propagation and normal direction of the $(q-1)$-th metasurface layer, and $\mathrm{d}^{q}_{m,m'}$ corresponds to the propagation distance. 

Assuming that the \ac{SIM} is placed very close to an $N_\mathrm{T}$-element TX \ac{ULA} with adjacent inter-element spacing $\lambda / 2$, we also define $\bm{\Gamma}_{1} \triangleq [\bm{\gamma}^1_1, \dots, \bm{\gamma}^1_{N_\mathrm{T}}] \in \mathbb{C}^{M \times N_\mathrm{T}}$, where $\bm{\gamma}^1_{n_\mathrm{T}} \in \mathbb{C}^{M \times 1}$ (with $n_\mathrm{T}=1,\ldots,N_\mathrm{T}$) represents the transmission vector from the $n_\mathrm{T}$-th transmit antenna to the innermost metasurface layer of the \ac{SIM}, whose $m$-th element $\gamma^1_{m,n_\mathrm{T}}$ is obtained by substituting $\epsilon^{q}_{m,m'}$ and $\mathrm{d}^{q}_{m,m'}$ in equation \eqref{eq:diffraction_coeff} with $\epsilon^1_{m,n_\mathrm{T}}$ and $\mathrm{d}^1_{m,n_\mathrm{T}}$, respectively. 
Concatenating the above, the overall $M\times N_\mathrm{T}$ propagation matrix from the \ac{TX} antenna elements to the meta-atoms of the $Q$-th \ac{SIM} layer can be expressed as
\begin{equation}
\bm{\Upsilon}_\mathrm{T}(\bm{\mathcal{Z}}) \triangleq \prod_{q=1}^Q \bm{\Psi}_{Q-q+1} \bm{\Gamma}_{Q-q+1},
\label{eq:transmit_SIM_full}
\end{equation}
where we have used the definition $\bm{\mathcal{Z}}\triangleq\{\bm{\Psi}_1,\ldots,\bm{\Psi}_Q\}$.

Similarly, a \ac{SIM} of $\tilde{\mathcal{Q}}$ layers of transmissive metasurfaces, each comprising $\tilde{M} \triangleq \tilde{M}_x \tilde{M}_z$ response-tunable meta-atoms with $\tilde{M}_x$ and $\tilde{M}_z$ being the number of meta-atoms in the $x$- and $z$-axis on each layer, respectively, placed very close to an $N_\mathrm{R}$-element RX \ac{ULA} with adjacent inter-element spacing $\lambda / 2$, results in the overall $N_\mathrm{R}\times \tilde{M}$ propagation matrix from the SIM to the RX antenna array described by
\begin{equation}
\bm{\Upsilon}_\mathrm{R}(\tilde{\bm{\mathcal{Z}}}) \triangleq \prod_{\tilde{q}=1}^{\tilde{Q}} \bm{\Xi}_{\tilde{q}} \bm{\Delta}_{\tilde{q}},
\label{eq:receive_SIM_full}
\end{equation}
where $\bm{\Xi}_{\tilde{q}} \in \mathbb{C}^{\tilde{M} \times \tilde{M}}$ $\forall \tilde{q} \in\{2,\ldots,\tilde{Q}\}$ is the transmission matrix between the $\tilde{q}$-th and $(\tilde{q}-1)$-th layer of the \ac{SIM}, whose elements are defined similar to equation \eqref{eq:diffraction_coeff} as
\begin{equation}
\label{eq:diffraction_coeff_rx}
\xi^{\tilde{q}}_{\tilde{m},\tilde{m}'} \triangleq \frac{\rho_r \cos\big(\tilde{\epsilon}^{\tilde{q}}_{\tilde{m},\tilde{m}'}\big)}{\tilde{\mathrm{d}}^{\tilde{q}}_{\tilde{m},\tilde{m}'}}  \bigg( \frac{1}{2\pi  \tilde{\mathrm{d}}^{\tilde{q}}_{\tilde{m},\tilde{m}'}} - \frac{\jmath}{\lambda} \bigg)  e^{\jmath2\pi \frac{\tilde{\mathrm{d}}^{\tilde{q}}_{\tilde{m},\tilde{m}'}}{\lambda}},
\end{equation}
where $\rho_r$ denotes the square measure occupied by each meta-atom in the \ac{RX}-\ac{SIM}, $\tilde{\epsilon}^{\tilde{q}}_{\tilde{m},\tilde{m}'}$ is the angle between the propagation and normal direction of the $(\tilde{q}-1)$-th metasurface layer and $\tilde{\mathrm{d}}^{\tilde{q}}_{\tilde{m},\tilde{m}'}$ corresponds to the propagation distance.

Subsequently, $\bm{\Xi}_{1} \triangleq [\bm{\xi}^1_1, \dots, \bm{\xi}^1_{N_\mathrm{R}}]\trans \in \mathbb{C}^{N_\mathrm{R} \times \tilde{M}}$ with $\bm{\xi}^1_{n_\mathrm{R}} \in \mathbb{C}^{\tilde{M} \times 1}$ ($n_\mathrm{R}=1,\ldots,N_\mathrm{R}$) denoting the transmission vector from the $n_\mathrm{R}$-th receive antenna to the innermost \ac{SIM} layer, whose $\tilde{m}$-th element $\xi^1_{\tilde{m},n_\mathrm{R}}$ with $\forall m \in \tilde{\mathcal{M}}\triangleq\{ 1, \dots, \tilde{M} \}$ is defined similarly to $\gamma^1_{m,n_\mathrm{T}}$, but using equation \eqref{eq:diffraction_coeff_rx}.
Finally, $\bm{\Delta}_{\tilde{q}}$, $\forall\tilde{q} \in \tilde{\mathcal{Q}}\triangleq\{1,\ldots,\tilde{Q}\}$ defined similar to equation \eqref{eq:diagona_shift_matrix_per_layer} including the effective tunable phase responses 
$\tilde{\zeta}^{\tilde{q}}_{\tilde{m}}\in [0,2\pi)$ $\forall \tilde{q} \in \tilde{\mathcal{Q}}$ and $\forall \tilde{m} \in \tilde{\mathcal{M}}$ with $\tilde{\bm{\mathcal{Z}}}\triangleq\{\bm{\Delta}_1,\ldots,\bm{\Delta}_{\tilde{Q}}\}$ is given by
\vspace{-1ex}
\begin{equation}
  \vspace{-1ex}
\bm{\Delta}_{\tilde{q}}  \triangleq \text{diag}\bigg( \!\Big[ e^{\jmath \tilde{\zeta}^{\tilde{q}}_{1}}, \dots, e^{\jmath \tilde{\zeta}^{\tilde{q}}_{\tilde{M}}} \Big]\! \bigg).
\label{eq:diagona_shift_matrix_per_layer_rx}
\end{equation}

\section{The Proposed MPDD MIMO Channel Model}
\label{MPDD_MIMO_Model}
\begin{figure*}[t!]
\centering
\includegraphics[width=0.9\textwidth]{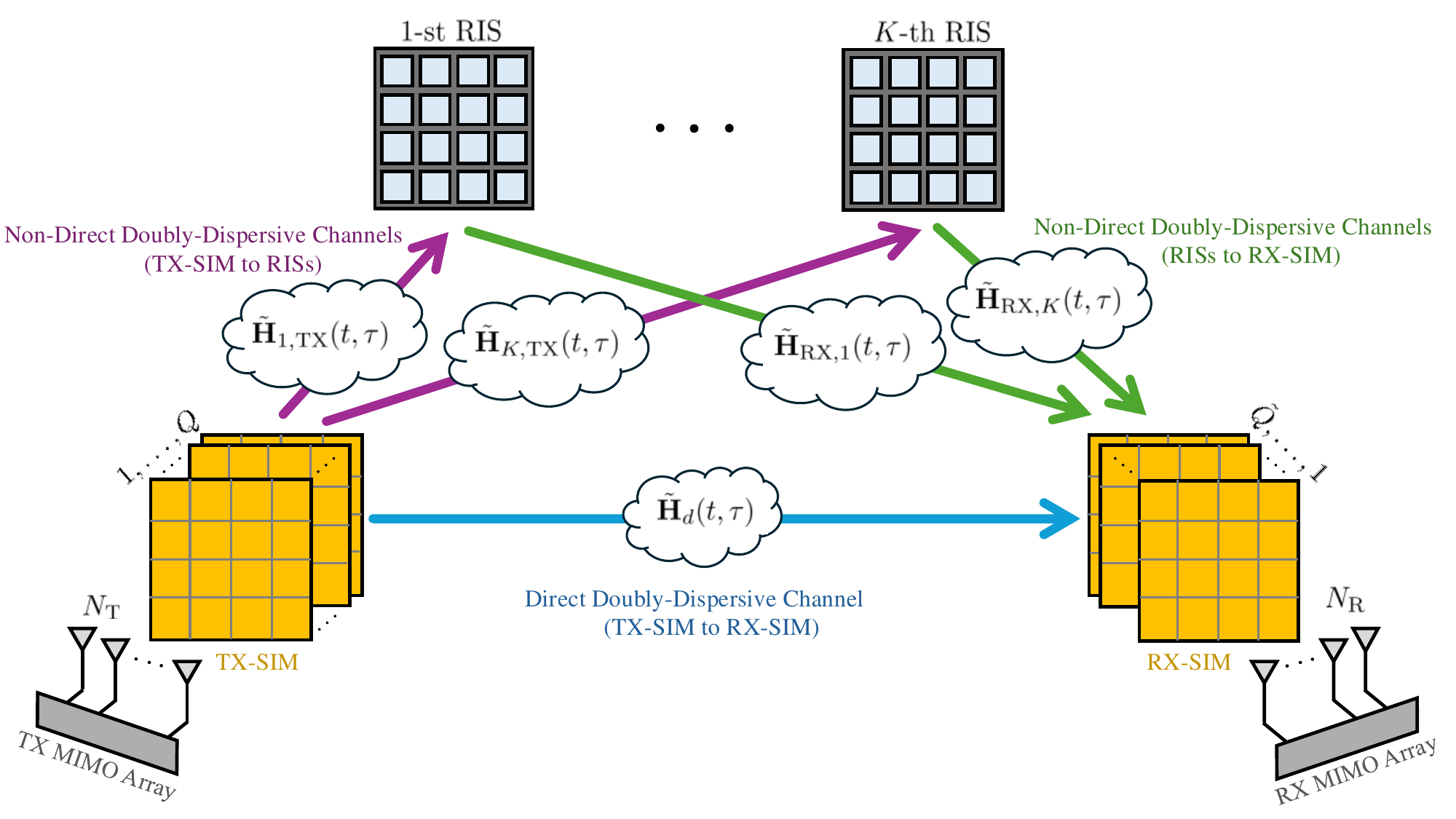}
\vspace{-1ex}
\caption{The considered MPDD \ac{MIMO} system for high-mobility scenarios, which includes two \acp{SIM}, one placed very close to the \ac{TX} and the other very close to the \ac{RX}, and $K$ \acp{RIS} within the wireless propagation environment of interest.}
\label{fig:system_model_SIM}
\vspace{-2ex}
\end{figure*}

Consider {a} point-to-point \ac{MIMO} system {as} illustrated in Fig.~\ref{fig:system_model_SIM}{, where} a transmitter equipped with an $N_\mathrm{T}$-element \ac{ULA} and a $Q$-layered \ac{SIM} {communicates with} a receiver equipped with a $\tilde{Q}$-layered \ac{SIM} and a \ac{ULA}-based front-end with $N_\mathrm{R}$ antennas\footnote{While the possibility to directly use \acp{SIM} as active radiating structures also exists \cite{BayraktarAsilimoar2024,MatemuTWC2025}, we here follow \cite{AnJSAC2023,Dardari_ARXIV_2024,ShiTWC2025,StutzOJCOMS2025,AnTWC2025} and assume these structures to be passive low-cost devices.}.
The \ac{TX} and \ac{RX} \acp{SIM}, {respectively denoted \ac{TX}-\ac{SIM} and \ac{RX}-\ac{SIM}}, are both {placed very close} to their respective antennas, as described in Section~\ref{sec:SIM_Modeling}.

{In addition, as seen from Fig.~\ref{fig:system_model_SIM},} the \ac{MIMO} system operates within a smart wireless environment comprising $K$ \acp{RIS} each consisting of $J \triangleq J_x J_z$ response-tunable reflective meta-atoms, {with} $J_x$ and $J_z$ denot{ing} the number of meta-atoms in the $x$- and $z$-axis, respectively. 
{Let $\phi^{k}_{j}\in [0,2\pi)$ denote the reflective phase response of each $j$-th ($j \in \{1,\ldots,J\}$) meta-atom contained in the $k$-th ($k \in \{1,\ldots,K\}$) \ac{RIS}.
Then,} the $J\times J$ phase configuration matrix $\bm{\Phi}_k$ of each $k$-th \ac{RIS} can be expressed similarly to eqs. \eqref{eq:diagona_shift_matrix_per_layer} and \eqref{eq:diagona_shift_matrix_per_layer_rx} as
\begin{equation}
\bm{\Phi}_k  \triangleq \text{diag}\bigg( \!\Big[ e^{\jmath\phi^{k}_{1}}, \dots, e^{\jmath\phi^{k}_{J}} \Big]\! \bigg).
\label{eq:diagona_shift_matrix_RIS}
\end{equation}

Then, {by parametrizing the configurations of the \ac{TX} and \ac{RX} \acp{SIM} (i.e., $\bm{\mathcal{Z}}$ and $\tilde{\bm{\mathcal{Z}}}$, respectively) as well as the \acp{RIS} (i.e., $\bm{\mathcal{F}}\triangleq\{\bm{\Phi}_1,\ldots,\bm{\Phi}_K\}$)}, the complex-valued $N_\mathrm{R} \times N_\mathrm{T}$ end-to-end \ac{MPDD} smart wireless \ac{MIMO} channel can be expressed as\footnote{The proposed MPDD \ac{MIMO} channel model holds also for cases where any of the \acp{SIM} or \acp{RIS} have a non-local structure \cite{NeriniTWC2024,WMA2024}.
{This is also true for cases caused by nonlinearities in the fundamental model, leading to changes in the Rayleigh-Sommerfeld diffraction coefficients.}}
\begin{equation}
\label{eq:prop_SIM_channel}
\!\!\!\!\mathbf{H}(\bm{\mathcal{Z}},\tilde{\bm{\mathcal{Z}}},\bm{\mathcal{F}},t,\tau) \!\triangleq\! \bm{\Upsilon}_\mathrm{R}(\tilde{\bm{\mathcal{Z}}})\mathbf{R}_\mathrm{RX}^{1/2} \tilde{\mathbf{H}}(\bm{\mathcal{F}},t,\tau) \mathbf{R}_\mathrm{TX}^{1/2}\bm{\Upsilon}_\mathrm{T}(\bm{\mathcal{Z}}),\!\!
\end{equation}    
where {the spatial correlation matrices at the outermost layer of the \ac{TX}-\ac{SIM} and \ac{RX}-\ac{SIM} are respectively defined as $\mathbf{R}_\mathrm{TX} \in \mathbb{C}^{M \times M}$ and $\mathbf{R}_\mathrm{RX} \in \mathbb{C}^{\tilde{M} \times \tilde{M}}$}.

{Note that this is a consequence of the sub-wavelength spacing of the adjacent meta-atoms on the $\mathcal{Q}$-th and $\tilde{\mathcal{Q}}$-th layers.} 
{In addition,} each $(m,m')$-th and $(\tilde{m},\tilde{m}')$-th element of $\mathbf{R}_\mathrm{TX}$ and $\mathbf{R}_\mathrm{RX}$ in equation \eqref{eq:prop_SIM_channel} are respectively defined as $[\mathbf{R}_\mathrm{TX}]_{m,m'} \triangleq \text{sinc}\left(2\mathrm{d}_{m,m'}/\lambda\right)$ and $[\mathbf{R}_\mathrm{RX}]_{\tilde{m},\tilde{m}'} \triangleq \text{sinc}\big(2\tilde{\mathrm{d}}_{\tilde{m},\tilde{m}'}/\lambda\big)$ {following \cite{AnJSAC2023}}.  

{Subsequently,} the complex-valued $\tilde{M} \times M$ \ac{DD} \acp{RIS}-parametrized channel matrix $\tilde{\mathbf{H}}(\bm{\mathcal{F}},t,\tau)$ is given by
\vspace{-1ex}
\begin{equation}
\label{eq:MIMO_TD_channel_general_SIM}
\tilde{\mathbf{H}}(\bm{\mathcal{F}},t,\tau) \!\triangleq\! \tilde{\mathbf{H}}_d(t,\tau) +\! \sum_{k=1}^{K} \tilde{\mathbf{H}}_{\mathrm{RX},k}(t,\tau) \bm{\Phi}_k \tilde{\mathbf{H}}_{k,\mathrm{TX}}(t,\tau),\!
\end{equation}
where $\tilde{\mathbf{H}}_d(t,\tau) \in \mathbb{C}^{\tilde{M} \times M}$ represents the direct $P$-path of the \ac{DD} \ac{MIMO} channel between the $\tilde{M}$-element $\tilde{\mathcal{Q}}$-th layer of the \ac{RX}-\ac{SIM} and the $M$-element $\mathcal{Q}$-th layer of the \ac{TX}-\ac{SIM}, {with its corresponding definition given by}
\vspace{-1ex}
\begin{eqnarray}
\label{eq:MIMO_TD_channel_general_SIM_LOS}
\tilde{\mathbf{H}}_d(t,\tau) \triangleq \sqrt{\tfrac{M\tilde{M}}{P}} \sum_{p=1}^P h_p e^{\jmath 2\pi \nu_p t} \delta\left(\tau-\tau_p\right)&&\\[-2ex] 
&&\hspace{-18ex}\times  \mathbf{b}_\mathrm{R}\left(\phi_p^{\rm in},\theta_p^{\rm in}\right) \mathbf{b}_\mathrm{T}\herm\left(\phi_p^{\rm out},\theta_p^{\rm out}\right),\nonumber
\end{eqnarray}
where $\mathbf{b}_\mathrm{T}(\cdot,\cdot) \in \mathbb{C}^{M \times 1}$ and $\mathbf{b}_\mathrm{R}(\cdot,\cdot) \in \mathbb{C}^{\tilde{M} \times 1}$ {defined in equation \eqref{eq:SIM_transmit_UPA_response vector}} are respectively the \ac{UPA} response vectors for the \ac{TX}-\ac{SIM} and \ac{RX}-\ac{SIM}, with $(\phi_p^{\rm in},\theta_p^{\rm in})$ and $(\phi_p^{\rm out},\theta_p^{\rm out})$ {denoting} the pairs of azimuth and elevation \acp{AoA} and \acp{AoD}, respectively, for each $p$-th signal propagation path {with} the complex channel gain $h_p$, with $p=\{1,\ldots,P\}$. 

In {addition}, $\tau_p \in [0,\tau_\text{max}]$ and $\nu_p \in [-\nu_\text{max},\nu_\text{max}]$ denote each $p$-th path's delay in seconds and Doppler shift in Hz, respectively.
Furthermore, {for future convenience, let us} define the \ac{UPA} response matrix associated with a given $p$-th path {as}
\begin{align}
\label{eq:Bmatrix}
\mathbf{B}_p &\triangleq  \mathbf{b}_\mathrm{R}\left(\phi_p^{\rm in},\theta_p^{\rm in}\right) \mathbf{b}_\mathrm{T}\herm\left(\phi_p^{\rm out},\theta_p^{\rm out}\right) .
\end{align}

Finally, for the sake of clarity we emphasize that the matrices $\tilde{\mathbf{H}}_{k,\mathrm{TX}}(t,\tau)\in\mathbb{C}^{J\times M}$ and $\tilde{\mathbf{H}}_{\mathrm{RX},k}(t,\tau)\in\mathbb{C}^{\tilde{M}\times J}$  in equation \eqref{eq:MIMO_TD_channel_general_SIM} can be expressed similarly to equation \eqref{eq:MIMO_TD_channel_general_SIM_LOS} as $\tilde{P}$- and $\bar{P}$-path \ac{DD} \ac{MIMO} channels, {given by}
\vspace{-1ex}
\begin{equation}
\label{eq:MIMO_TD_channel_general_SIM_nLOS_TX}
\tilde{\mathbf{H}}_{k,\mathrm{TX}}(t,\tau) \triangleq \sqrt{\tfrac{JM}{\tilde{P}}} \sum_{\tilde{p}=1}^{\tilde{P}} h_{\tilde{p},k} e^{\jmath2\pi \nu_{\tilde{p},k} t} \delta( \tau - \tau_{\tilde{p},k}) \mathbf{B}_{\tilde{p},k},
%
%
\end{equation}
and
\vspace{-1ex}
\begin{equation}
\label{eq:MIMO_TD_channel_general_SIM_nLOS_RX}
\tilde{\mathbf{H}}_{\mathrm{RX},k}(t,\tau) \triangleq \sqrt{\tfrac{J\tilde{M}}{\bar{P}}}  \sum_{\bar{p}=1}^{\bar{P}} h_{\bar{p},k} e^{\jmath 2\pi \nu_{\bar{p},k} t} \delta( \tau - \tau_{\bar{p},k}) \mathbf{B}_{\bar{p},k}.
%
%
\end{equation}

The notation $h_{\tilde{p},k}$ as well as $(\phi_{\tilde{p},k}^{\rm in},\theta_{\tilde{p},k}^{\rm in})$ and $(\phi_{\tilde{p},k}^{\rm out},\theta_{\tilde{p},k}^{\rm out})$ {will be used henceforth to denote} the complex channel gain and pairs of azimuth and elevation \acp{AoA} and \acp{AoD}, respectively, for each $\tilde{p}$-th signal propagation path between the \ac{TX}-\ac{SIM} and each $k$-th \ac{RIS}, with $\tilde{p}=\{1,\ldots,\tilde{P}\}$. 
Similarly, $h_{\bar{p},k}$ as well as $(\phi_{\bar{p},k}^{\rm in},\theta_{\bar{p},k}^{\rm in})$ and $(\phi_{\bar{p},k}^{\rm out},\theta_{\bar{p},k}^{\rm out})$ indicate the complex channel gain and pairs of azimuth and elevation \acp{AoA} and \acp{AoD}, respectively, for each $\bar{p}$-th path between each $k$-th \ac{RIS} and the \ac{RX}-\ac{SIM}, with $\bar{p}=\{1,\ldots,\bar{P}\}$.
The delays and  Doppler shifts of the latter paths are denoted by $\tau_{\tilde{p},k}$, $\nu_{\tilde{p},k}$ and $\tau_{\bar{p},k}$, $\nu_{\bar{p},k}$, respectively, having similar bounds to $\tau_p$ and $\nu_p$.

\begin{remark}[Special Cases]
Removing the second term with the $K$ summations in equation \eqref{eq:MIMO_TD_channel_general_SIM} simplifies the full model to the \ac{MIMO} channel case including a \ac{TX} and \ac{RX} \ac{SIM} with no \acp{RIS} present.
In addition, removing any of the factors $\bm{\Upsilon}_\mathrm{R}(\tilde{\bm{\mathcal{Z}}})\mathbf{R}_\mathrm{RX}^{1/2}$ or $\mathbf{R}_\mathrm{TX}^{1/2}\bm{\Upsilon}_\mathrm{T}(\bm{\mathcal{Z}})$ in equation \eqref{eq:prop_SIM_channel}, implies a \ac{MIMO} system lacking a \ac{RX}-\ac{SIM} or a \ac{TX}-\ac{SIM}, respectively. 
In the second of the latter two cases, $\tilde{\mathbf{H}}(\bm{\mathcal{F}},t,\tau)$ appearing in equations \eqref{eq:prop_SIM_channel} and \eqref{eq:MIMO_TD_channel_general_SIM} models signal propagation directly from the elements of the \ac{TX} \ac{ULA} (see the \ac{ULA} response vector in equation \eqref{eq:ULA_response_vector_transmit}) to the \ac{RX}-\ac{SIM}, via the $K$ \acp{RIS}.
In this case, the size of $\tilde{\mathbf{H}}(\bm{\mathcal{F}},t,\tau)$ becomes $\tilde{M}\times N_\mathrm{T}$. 
Alternatively, when only the \ac{RX}-\ac{SIM} is missing, $\tilde{\mathbf{H}}(\bm{\mathcal{F}},t,\tau)$ will be $ N_\mathrm{R}\times M$. 
Finally, when none of the \acp{SIM} are considered, $\mathbf{H}(\bm{\mathcal{Z}},\tilde{\bm{\mathcal{Z}}},\bm{\mathcal{F}},t,\tau)\equiv\tilde{\mathbf{H}}(\bm{\mathcal{F}},t,\tau)$ representing the \acp{RIS}-empowered $N_\mathrm{R}\times N_\mathrm{T}$ \ac{DD} \ac{MIMO} channel. 
In the absence of \acp{RIS} and for a single-antenna \ac{TX} and \ac{RX} (i.e., $N_\mathrm{T}=N_\mathrm{R}=1$), the latter channel model reduces to the model described in \cite{Rou_SPM_2024}. 
\end{remark}

\renewcommand{\arraystretch}{1.25}
\setlength{\tabcolsep}{1.2pt}
\begin{table}[H]

\caption{Variable Notation and Descriptions.}
\centering
\begin{tabular}{|c|c|}
\hline
\textbf{Variable} & \textbf{Description} \\
\hline
$N_\mathrm{T}$, $N_\mathrm{R}$ & Number of \ac{TX} and \ac{RX} antennas \\
\hline
$Q$, $\tilde{Q}$ & Number of \ac{TX}- and \ac{RX}-\ac{SIM} layers \\
\hline
$M$, $\tilde{M}$ & Number of \ac{TX}-\ac{SIM} and \ac{RX}-\ac{SIM} meta-atoms per layer \\
\hline
$K$ & Number of \acp{RIS} \\
\hline
$J$ & Number of meta-atoms on each \ac{RIS} \\
\hline
$\bm{\Phi}_k$ & Phase configuration matrix of the $k$-th \ac{RIS} \\
\hline
$\bm{\mathcal{Z}}$, $\tilde{\bm{\mathcal{Z}}}$, $\bm{\mathcal{F}}$ & Phase sets of the \ac{TX}-\ac{SIM}, \ac{RX}-\ac{SIM} and \acp{RIS} \\
\hline
$\bm{\Upsilon}_\mathrm{T}(\tilde{\bm{\mathcal{Z}}})$, $\bm{\Upsilon}_\mathrm{R}(\tilde{\bm{\mathcal{Z}}})$ & \ac{TX}-\ac{SIM} and \ac{RX}-\ac{SIM} transfer functions \\
\hline
$\mathbf{R}_\mathrm{TX}^{1/2}$, $\mathbf{R}_\mathrm{RX}^{1/2}$ & \ac{TX}-\ac{SIM} and \ac{RX}-\ac{SIM} spatial correlation matrices  \\
\hline
$\mathbf{b}_\mathrm{T}(\cdot,\cdot)$, $\mathbf{b}_\mathrm{R}(\cdot,\cdot)$ & \ac{TX} and \ac{RX} \ac{UPA} response vectors \\
\hline
$\tilde{\mathbf{H}}_d(t,\tau)$ & Direct \ac{DD} \ac{MIMO} channel between the \acp{SIM} \\
\hline
$\tilde{\mathbf{H}}_{k,\mathrm{TX}}(t,\tau)$ & \ac{DD} \ac{MIMO} channel between the \ac{TX}-\ac{SIM} and \acp{RIS}  \\
\hline
$\tilde{\mathbf{H}}_{\mathrm{RX},k}(t,\tau)$ & \ac{DD} \ac{MIMO} channel between the \acp{RIS} and the \ac{RX}-\ac{SIM} \\
\hline
$\tilde{\mathbf{H}}(\bm{\mathcal{F}},t,\tau)$ & \ac{DD} \acp{RIS}-parametrized channel matrix \\
\hline
$\mathbf{H}(\bm{\mathcal{Z}},\tilde{\bm{\mathcal{Z}}},\bm{\mathcal{F}},t,\tau)$ & End-to-end \ac{MPDD} smart wireless \ac{MIMO} channel \\
\hline
\end{tabular}
\label{tab:example}
\vspace{-2ex}
\end{table}

{

Notice that by modelling the \ac{MPDD} channel model in equation \eqref{eq:prop_SIM_channel} as a function of the phase configurations of the \ac{TX}-\ac{SIM}, \ac{RX}-\ac{SIM} and \acp{RIS}, as well as the spatial correlation matrices at the outermost layers of the \acp{SIM}, we have effectively generalized the model in \cite{Rou_SPM_2024} for reconfigurable electromagnetic environments, enabling the design of the relevant detection, estimation and resource allocation algorithms.

Table \ref{tab:example} summarizes the notation used in this section in conjunction with the corresponding description.}

\begin{figure*}[t!]
\setcounter{equation}{17}
\normalsize
\begin{align}
\mathbf{r}[n]\! =\! \sum_{\ell=0}^\infty \bigg[ \bigg(&\sum_{k=1}^K  \sum_{\bar{p}=1}^{\bar{P}} \sum_{\tilde{p}=1}^{\tilde{P}} \overbrace{J \sqrt{\tfrac{\tilde{M} M}{\bar{P}\tilde{P}}} h_{\bar{p},k} h_{\tilde{p},k}  \mathbf{U}\herm \bm{\Upsilon}_\mathrm{R}(\tilde{\bm{\mathcal{Z}}})  \mathbf{R}_\mathrm{RX}^{1/2} \mathbf{B}_{\bar{p},k}
%
%
\bm{\Phi}_k \mathbf{B}_{\tilde{p},k}
%
%
\mathbf{R}_\mathrm{TX}^{1/2}  \bm{\Upsilon}_\mathrm{T}(\bm{\mathcal{Z}}) \mathbf{V}}^{\triangleq\check{\mathbf{H}}_{k,\bar{p},\tilde{p}}^\mathrm{RIS}(\bm{\mathcal{Z}},\tilde{\bm{\mathcal{Z}}},\bm{\Phi}_k,\mathbf{V},\mathbf{U}) \in \mathbb{C}^{d_s \times d_s}} e^{\jmath 2\pi\frac{n}{N} \overbrace{(f_{\bar{p},k}\! +\! f_{\tilde{p},k})}^{\triangleq\hat{f}_{k,\bar{p},\tilde{p}}}}  \delta[ \ell\! -\! \overbrace{(\ell_{\bar{p},k}\! +\! \ell_{\tilde{p},k})}^{\triangleq\hat{\ell}_{k,\bar{p},\tilde{p}}} ] \nonumber \\
& + \sum_{p=1}^P \underbrace{\sqrt{\tfrac{M \tilde{M}}{P}} h_p \mathbf{U}\herm  \bm{\Upsilon}_\mathrm{R}(\tilde{\bm{\mathcal{Z}}})  \mathbf{R}_\mathrm{RX}^{1/2} \mathbf{B}_p
%
%
\mathbf{R}_\mathrm{TX}^{1/2}  \bm{\Upsilon}_\mathrm{T}(\bm{\mathcal{Z}}) \mathbf{V}}_{\triangleq\check{\mathbf{H}}_p^d(\bm{\mathcal{Z}},\tilde{\bm{\mathcal{Z}}},\mathbf{V},\mathbf{U}) \in \mathbb{C}^{d_s \times d_s}}  e^{\jmath 2\pi f_p \frac{n}{N}}  \delta[ \ell\! -\! \ell_p ] \bigg)  \mathbf{s}[n\! -\! \ell] \bigg]\! + \!\mathbf{w}[n]
\label{eq:sampled_TD}
\end{align}
\hspace{30ex} \hrulefill \hspace{30ex}
\setcounter{equation}{18}
\begin{align}
\label{eq:vectorized_TD_IO}
\mathbf{r}_v = \sum_{u=1}^{d_s} \bigg( \overbrace{\sum_{p=1}^P \check{h}_{p,v,u}^d  \underbrace{\mathbf{\Theta}_p  \mathbf{\Omega}^{f_p}  \mathbf{\Pi}^{\ell_p}}_{\triangleq\mathbf{G}_p \in \mathbb{C}^{N \times N}}}^{\triangleq\bar{\mathbf{H}}_{v,u}^d(\bm{\mathcal{Z}},\tilde{\bm{\mathcal{Z}}},\mathbf{V},\mathbf{U}) \in \mathbb{C}^{N \times N}} + \overbrace{\sum_{k=1}^{K}  \sum_{\bar{p}=1}^{\bar{P}} \sum_{\tilde{p}=1}^{\tilde{P}} \check{h}_{k,\bar{p},\tilde{p},v,u}^\mathrm{RIS} \overbrace{\mathbf{\Theta}_{k,\bar{p},\tilde{p}}  \mathbf{\Omega}^{\hat{f}_{k,\bar{p},\tilde{p}}}  \mathbf{\Pi}^{\hat{\ell}_{k,\bar{p},\tilde{p}}}}^{\triangleq\mathbf{G}_{k,\bar{p},\tilde{p}} \in \mathbb{C}^{N \times N}}}^{\triangleq\bar{\mathbf{H}}_{v,u}^\mathrm{RIS}(\bm{\mathcal{Z}},\tilde{\bm{\mathcal{Z}}},\bm{\mathcal{F}},\mathbf{V},\mathbf{U}) \in \mathbb{C}^{N \times N}}   \bigg)  \mathbf{s}_u \!+\! \mathbf{w}_v \\[-3ex]
&\hspace{-25ex}= \sum_{u=1}^{d_s} \!\underbrace{\Big(\bar{\mathbf{H}}_{v,u}^d(\bm{\mathcal{Z}},\tilde{\bm{\mathcal{Z}}},\mathbf{V},\mathbf{U}) \!+\! \bar{\mathbf{H}}_{v,u}^\mathrm{RIS}(\bm{\mathcal{Z}},\tilde{\bm{\mathcal{Z}}},\bm{\mathcal{F}},\mathbf{V},\mathbf{U})\Big)}_{\triangleq\bar{\mathbf{H}}_{v,u}^\mathrm{tot}(\bm{\mathcal{Z}},\tilde{\bm{\mathcal{Z}}},\bm{\mathcal{F}},\mathbf{V},\mathbf{U}) \in \mathbb{C}^{N \times N}}  \!\mathbf{s}_u \!+\! \mathbf{w}_v\nonumber
\end{align}
\hspace{30ex} \hrulefill \hspace{30ex}
\setcounter{equation}{19}
\normalsize
\begin{equation}
\label{eq:diagonal_CP_matrix_def}
\mathbf{\Theta}_p \triangleq \text{diag}\Big( [ \underbrace{e^{-\jmath 2\pi {\phi_\mathrm{CP}(\ell_p)}}, e^{-\jmath 2\pi {\phi_\mathrm{CP}(\ell_p - 1)}}, \dots, e^{-\jmath 2\pi {\phi_\mathrm{CP}(2)}}, e^{-\jmath 2\pi {\phi_\mathrm{CP}(1)}}}_{\ell_p \; \text{terms}}, \underbrace{1, 1, \dots, 1, 1}_{N - \ell_p \; \text{ones}}] \Big) \in \mathbb{C}^{N \times N}
\end{equation}
\hspace{30ex} \hrulefill \hspace{30ex}
\setcounter{equation}{20}
\begin{equation}
\label{eq:diagonal_Doppler_matrix_def}
\boldsymbol{\Omega} \triangleq \text{diag}\Big([1,e^{-\jmath 2\pi /N},\dots,e^{-\jmath 2\pi (N-2) /N}, e^{-\jmath 2\pi (N-1) /N}]\Big) \in \mathbb{C}^{N \times N}
\end{equation}

\setcounter{equation}{16}
\hrulefill
\vspace{-3ex}
\end{figure*}

\section{I/O Relationships for OFDM, OTFS, and AFDM}
\label{IO_Model}

In this section, we present various expressions for the \ac{TD} received signal corresponding to waveforms (such as \ac{OFDM}, \ac{OTFS}, and \ac{AFDM}) under the proposed \ac{MPDD} \ac{MIMO} channel model, effectively extending the work in \cite{Rou_SPM_2024}.

\subsection{Arbitrarily Modulated Signals}

Suppose that the point-to-point \ac{MIMO} system of Fig.~\ref{fig:system_model_SIM} deploys fully digital beamformers at its $N_\mathrm{T}$-element \ac{TX} and $N_\mathrm{R}$-element \ac{RX}; the extension to hybrid analog and digital \ac{BF} is straightforward and is left for future investigation.

Let $\mathbf{V} \in \mathbb{C}^{N_\mathrm{T} \times d_s}$ and $\mathbf{U} \in \mathbb{C}^{N_\mathrm{R} \times d_s}$ represent the transmit and receive digital beamformers, respectively, where $d_s \triangleq \text{min}(N_\mathrm{T},N_\mathrm{R})$ indicates the number of independent data streams to be communicated per coherent channel block. 
In what follows, the complex-valued $d_s$-element vector $\mathbf{s}(t)$ represents the power-limited transmit signal of any modulation (e.g., \ac{OFDM}, \ac{OTFS}, and \ac{AFDM}) in the \ac{TD}. 

Correspondingly, the $d_s$-element baseband received signal at a time instant $t$ (after the digital combiner) through the \ac{MPDD} \ac{MIMO} channel can be mathematically expressed as 
\begin{eqnarray}
\label{eq:TD_I/O_relationship}
\mathbf{r}(t) \triangleq  \mathbf{U}\herm \mathbf{H}(\mathbf{\bm{Z}},\tilde{\bm{\mathcal{Z}}},\bm{\mathcal{F}},t,\tau) * \mathbf{V} \mathbf{s}(t) + \mathbf{w}(t) = &&\\
&&\hspace{-46ex} \int\limits_{-\infty}^\infty \!\!\! \mathbf{U}\herm \bm{\Upsilon}_\mathrm{R}(\tilde{\bm{\mathcal{Z}}}) \mathbf{R}_\mathrm{RX}^{1/2} \bigg[ \tilde{\mathbf{H}}_d(t,\!\tau)\! +\!\!\! \sum_{k=1}^{K} \tilde{\mathbf{H}}_{\mathrm{RX},k}(t,\!\tau) \bm{\Phi}_k \tilde{\mathbf{H}}_{k,\mathrm{TX}}(t,\!\tau) \!\bigg]  \nonumber \\
&&\hspace{-24ex} \times \mathbf{R}_\mathrm{TX}^{1/2} \bm{\Upsilon}_\mathrm{T}(\bm{\mathcal{Z}}) \mathbf{V} \mathbf{s}(t - \tau) d\tau + \mathbf{w}(t),\nonumber
\end{eqnarray}
where $\mathbf{w}(t) \triangleq \mathbf{U}\herm \mathbf{n}(t) \in \mathbb{C}^{d_s \times 1}$ and $\mathbf{n}(t) \in \mathbb{C}^{N_\mathrm{R} \times 1}$ denotes the \ac{AWGN} vector at the RX side with spatially and temporally uncorrelated elements, each with zero mean and variance $\sigma_n^2$. 

Let $\mathbf{r}[n] \in \mathbb{C}^{d_s \times 1}$ and $\mathbf{s}[n] \in \mathbb{C}^{d_s \times 1}$, with $n \in \{ 0,\dots,N-1 \}$, be the finite sequences obtained after respectively sampling $\mathbf{r}(t)$ and $\mathbf{s}(t)$ at a sufficiently high sampling rate $F_S \triangleq \frac{1}{T_S}$ in Hz within a total bandwidth $B$. 
The discrete-time equivalent of the received signal in equation \eqref{eq:TD_I/O_relationship} can be obtained as portrayed (at the top of the {next} page) in equation \eqref{eq:sampled_TD}, where $\ell$ indicates the normalized discrete delay index, while $f_p \triangleq \frac{N\nu_p}{F_s}$ and $\ell_p \triangleq \frac{\tau_p}{T_s}$ are the normalized Doppler shift and the associated normalized discrete delay index of each $p$-th path propagation path between the \ac{TX}-\ac{SIM} and \ac{RX}-\ac{SIM}, respectively, with the definitions of $f_{\bar{p},k}$, $f_{\tilde{p},k}$, $\ell_{\bar{p},k}$, and $\ell_{\tilde{p},k}$ are similar for the respective channel paths. 

By taking into account a \ac{CP} of length $N_\mathrm{CP}$ and utilizing the circular convolution, the $N$-element discrete-time received signal in equation \eqref{eq:sampled_TD} can be re-expressed \cite{Rou_SPM_2024} as in equation \eqref{eq:vectorized_TD_IO}, where for notational simplicity we omit (also henceforth) the discrete-time index, which is implied.
In the latter equation, the scalars
$\check{h}_{p,v,u}^d$ and $\check{h}_{k,\bar{p},\tilde{p},v,u}^\mathrm{RIS}$, with $(v,u) = \{ 1,\dots,d_s \}$,  are respectively the $(v,u)$-th elements of the matrices $\check{\mathbf{H}}_p^d(\bm{\mathcal{Z}},\tilde{\bm{\mathcal{Z}}},\mathbf{V},\mathbf{U})$ and $\check{\mathbf{H}}_{k,\bar{p},\tilde{p}}^\mathrm{RIS}(\bm{\mathcal{Z}},\tilde{\bm{\mathcal{Z}}},\bm{\Phi}_k,\mathbf{V},\mathbf{U})$, both implicitly defined in equation \eqref{eq:sampled_TD}.
In addition, $\mathbf{s}_u \triangleq [s_u[0],\ldots,s_u[N-1]] \in \mathbb{C}^{N \times 1}$ and $\mathbf{w}_v \triangleq [w_v[0],\ldots,w_v[N-1]] \in \mathbb{C}^{N \times 1}$ are the transmit signal and \ac{AWGN}  vectors for the $u$-th and $v$-th stream, respectively.

In turn, each diagonal matrix $\mathbf{\Theta}_p \in \mathbb{C}^{N \times N}$ defined inside equation \eqref{eq:diagonal_CP_matrix_def} captures the effect of the \ac{CP} onto the $p$-th channel path, with $\phi_\mathrm{CP}(n)$ being a function of the sample index $n \in \{ 0,\ldots,N-1 \}$, representing a phase that depends on the specific waveform used.
In addition, the diagonal matrix $\boldsymbol{\Omega} \in \mathbb{C}^{N \times N}$ defined in equation \eqref{eq:diagonal_Doppler_matrix_def} contains $N$ complex roots of the unity, while $\mathbf{\Pi}\in \{0,1\}^{N \times N}$ is the forward cyclic shift matrix with elements defined as\footnotemark
\vspace{-1ex}
\setcounter{equation}{21}
\begin{equation}
\label{eq:PiMatrix}
\pi_{i,j} \triangleq \delta_{i,j+1} + \delta_{i,j-(N-1)}\,\;\; \delta _{ij} \triangleq
\begin{cases}
0 & \text{if }i\neq j\\
1 & \text{if }i=j
\end{cases}.
\vspace{-0.25ex}
\end{equation}

Leveraging the Kronecker product to concatenate all $d_s$ $\mathbf{r}_v$ vectors in equation \eqref{eq:vectorized_TD_IO}, the following $N d_s$-element vector for the overall received signal in the \ac{TD}, considering an arbitrary modulated transmit signal, is obtained as
\begin{equation}
\mathbf{r}_\mathrm{TD} = \bar{\mathbf{H}}(\bm{\mathcal{Z}},\tilde{\bm{\mathcal{Z}}},\bm{\mathcal{F}},\mathbf{V},\mathbf{U})  \mathbf{s}_\mathrm{TD} + \bar{\mathbf{w}}_\mathrm{TD},
\label{eq:vectorized_TD_IO_kron}
\end{equation}
where $\bar{\mathbf{H}}(\bm{\mathcal{Z}},\tilde{\bm{\mathcal{Z}}},\bm{\mathcal{F}},\mathbf{V},\mathbf{U})\in \mathbb{C}^{N d_s \times N d_s}$ explicitly highlights the dependence of the \ac{TD} transfer function of the considered point-to-point \ac{MIMO} system on the \ac{TX} and \ac{RX} \acp{SIM}, the $K$ \acp{RIS} of the programmable smart wireless propagation environment, and the digital \ac{TX} and \ac{RX} beamformers, and is mathematically defined as
\begin{eqnarray}
\label{eq:H_bar}
\bar{\mathbf{H}}(\bm{\mathcal{Z}},\tilde{\bm{\mathcal{Z}}},\bm{\mathcal{F}},\mathbf{V},\mathbf{U})\triangleq\sum_{p=1}^P (\check{\mathbf{H}}_p^d(\bm{\mathcal{Z}},\tilde{\bm{\mathcal{Z}}},\mathbf{V},\mathbf{U}) \otimes \mathbf{G}_p)&& \\[-1ex]
&&\hspace{-45ex}+ \sum_{k=1}^{K}  \sum_{\bar{p}=1}^{\bar{P}} \sum_{\tilde{p}=1}^{\tilde{P}} (\check{\mathbf{H}}_{k,\bar{p},\tilde{p}}^\mathrm{RIS}(\bm{\mathcal{Z}},\tilde{\bm{\mathcal{Z}}},\bm{\Phi}_k,\mathbf{V},\mathbf{U}) \otimes \mathbf{G}_{k,\bar{p},\tilde{p}}),\nonumber
\end{eqnarray}
with the $N d_s$-element vectors $\mathbf{s}_\mathrm{TD}$ and $\bar{\mathbf{w}}_\mathrm{TD}$ resulting from the concatenation of $\mathbf{s}_u$'s and $\mathbf{w}_v$'s in equation \eqref{eq:vectorized_TD_IO}, respectively. 

%
\setcounter{equation}{23}

\footnotetext{The matrix $\mathbf{\Pi}$ is defined such that $\bm{A}\mathbf{\Pi}^{\ell_p}$, with $\ell_p\in\mathbb{N}$, is a cyclic left-shifted version of $\bm{A}$, $i.e.$, the first $\ell_p$ columns of $\bm{A}$ are moved to the positions of the last $\ell_p$ columns. It can be also seen that $\mathbf{\Pi}^0$ is the $N \times N$ identity matrix, yielding $\bm{A}\mathbf{\Pi}^0=\bm{A}$.}

For notational simplicity, the matrices $\check{\mathbf{H}}_p^d$ and $\check{\mathbf{H}}_{k,\bar{p},\tilde{p}}^\mathrm{RIS}$ appearing in equation \eqref{eq:H_bar} will hereafter be expressed without explicitly indicating their dependence on the \ac{TX}/\ac{RX} \ac{BF} and \acp{SIM}/\acp{RIS} parameters.

{

\begin{remark}[\ac{SIM} Parametrization vs. Hybrid \ac{BF}]
As seen from equation \eqref{eq:H_bar}, the final \ac{TD} channel matrix is a function of \ul{both} the classical \ac{TX}/\ac{RX} \ac{BF} and the \acp{SIM}/\acp{RIS} parameters.
While the \ac{TX}/\ac{RX} \ac{BF} and \acp{SIM}/\acp{RIS} parameters could be jointly optimized leading to a complicated formulation, it is worth noting that the \ac{TX}/\ac{RX} \ac{BF} can be designed independently of the \acp{SIM}/\acp{RIS} parameters, depending on the specific application.
Therefore, \acp{SIM} and \acp{RIS} can be considered augmentations to the \ac{TX}/\ac{RX} \ac{BF} design, which can be optimized separately and will be discussed further in Section~\ref{sec:joint_optimization}.
\end{remark}

}

%
%
\begin{figure*}[t!]
\setcounter{equation}{38}
\normalsize
\begin{equation}
\label{eq:AFDM_diagonal_CP_matrix_def}
\bm{\varTheta}_p \triangleq \text{diag}\bigg( [ \underbrace{e^{-\jmath 2\pi {c_1} (N^2-2N\ell_p)}, e^{-\jmath 2\pi {c_1} (N^2-2N(\ell_p-1))}, \dots, e^{-\jmath 2\pi {c_1} (N^2-2N)}}_{\ell_p \; \text{terms}}, \underbrace{1, 1, \dots, 1, 1}_{N - \ell_p \; \text{ones}}] \bigg) \in \mathbb{C}^{N \times N}
\vspace{-1ex}
\end{equation}
\hrulefill
\vspace{-3ex}
\end{figure*}

\setcounter{equation}{24}

\vspace{-2ex}
\subsection{OFDM Signaling}

Let $\mathcal{C}$ denote an arbitrary complex constellation set of cardinality $D$ and average energy $E_\mathrm{S}$, which is associated with a given digital modulation scheme (e.g., \ac{QAM}). In \ac{OFDM}, multiple information vectors $\mathbf{x}_u \in \mathcal{C}^{N\times 1}$ with $u = \{ 1,\dots,d_s \}$, containing a total of $Nd_s$ symbols, are modulated into the following transmit signal as
\begin{equation}
\label{eq:OFDM_modulation}
\mathbf{s}^{(\text{OFDM})}_u \triangleq \mathbf{F}_N\herm  \mathbf{x}_u \in \mathbb{C}^{N \times 1},
\end{equation}
where $\mathbf{F}_N$ denotes the $N$-point normalized \ac{DFT} matrix. 

After undergoing circular convolution with the \ac{DD} channel and using a formulation similar to equation \eqref{eq:vectorized_TD_IO_kron}, the corresponding $Nd_s$-element discrete-time received \ac{OFDM} signal can be written as
\begin{equation}
\label{eq:TD_OFDM_input_output}
\mathbf{r}_\text{OFDM} \triangleq \bar{\mathbf{H}}(\bm{\mathcal{Z}},\tilde{\bm{\mathcal{Z}}},\bm{\mathcal{F}},\mathbf{V},\mathbf{U})  \mathbf{s}_\text{OFDM} + \bar{\mathbf{w}}_\mathrm{TD},
\end{equation}
where the $Nd_s$-element vectors are defined as
\begin{equation}
\label{eq:OFDM_stacked_s}
\mathbf{s}_\text{OFDM} \triangleq 
\begin{bmatrix}
\mathbf{s}^{(\text{OFDM})}_1 \\[-1ex]
\vdots \\
\mathbf{s}^{(\text{OFDM})}_{d_s}
\end{bmatrix},\,\,
\mathbf{r}_\text{OFDM} \triangleq 
\begin{bmatrix}
\mathbf{r}^{(\text{OFDM})}_1 \\[-1ex]
\vdots \\
\mathbf{r}^{(\text{OFDM})}_{d_s}
\end{bmatrix}.
\end{equation}

At the \ac{RX} side, applying \ac{OFDM} demodulation yields
\begin{equation}
\label{eq:OFDM_demodulation}
\mathbf{y}^{(\text{OFDM})}_v \triangleq \mathbf{F}_N  \mathbf{r}^{(\text{OFDM})}_v \in \mathbb{C}^{N \times 1},
\end{equation}
yielding the corresponding $Nd_s$-element discrete-time signal
\begin{equation}
\label{eq:OFDM_input_output}
\mathbf{y}_\text{OFDM} = \bar{\mathbf{H}}_\text{OFDM}(\bm{\mathcal{Z}},\tilde{\bm{\mathcal{Z}}},\bm{\mathcal{F}},\mathbf{V},\mathbf{U})  \mathbf{x} + \bar{\mathbf{w}}_\text{OFDM}, 
\end{equation}

\noindent where $\bar{\mathbf{w}}_\text{OFDM} \in \mathbb{C}^{Nd_s \times 1}$ is an equivalent \ac{AWGN} with the same statistics as $\bar{\mathbf{w}}_\mathrm{TD}$, and $\bar{\mathbf{H}}_\text{OFDM}(\bm{\mathcal{Z}},\tilde{\bm{\mathcal{Z}}},\bm{\mathcal{F}},\mathbf{V},\mathbf{U}) \in \mathbb{C}^{Nd_s \times Nd_s}$ represents the effective \ac{OFDM} channel defined similar to $\bar{\mathbf{H}}(\bm{\mathcal{Z}},\tilde{\bm{\mathcal{Z}}},\bm{\mathcal{F}},\mathbf{V},\mathbf{U})$ in equation\eqref{eq:vectorized_TD_IO_kron} as
\begin{eqnarray}
\label{eq:OFDM_effective_channel}
\bar{\mathbf{H}}_\text{OFDM}
\triangleq \sum_{p=1}^P \check{\mathbf{H}}_p^d \otimes \overbrace{( \mathbf{F}_N \mathbf{G}_p  \mathbf{F}_N\herm)}^{\triangleq\mathbf{G}_p^\text{OFDM} \in \mathbb{C}^{N \times N}}&& \\[-4ex]
&&\hspace{-30ex}\quad\quad\quad\quad+\sum_{k=1}^{K}  \sum_{\bar{p}=1}^{\bar{P}} \sum_{\tilde{p}=1}^{\tilde{P}} \check{\mathbf{H}}_{k,\bar{p},\tilde{p}}^\mathrm{RIS} \otimes \overbrace{( \mathbf{F}_N \mathbf{G}_{k,\bar{p},\tilde{p}}  \mathbf{F}_N\herm)}^{\triangleq\mathbf{G}_{k,\bar{p},\tilde{p}}^\text{OFDM} \in \mathbb{C}^{N \times N}}\nonumber \\[-1ex]
&&\hspace{-28ex}= \sum_{p=1}^P \check{\mathbf{H}}_p^d \otimes {\mathbf{G}_p^\text{OFDM}} + \sum_{k=1}^{K}  \sum_{\bar{p}=1}^{\bar{P}} \sum_{\tilde{p}=1}^{\tilde{P}} \check{\mathbf{H}}_{k,\bar{p},\tilde{p}}^\mathrm{RIS} \otimes {\mathbf{G}_{k,\bar{p},\tilde{p}}^\text{OFDM}}.\nonumber
\end{eqnarray}

Notice that for the \ac{OFDM} case, the \ac{CP} phase matrices $\mathbf{\Theta}_p$'s appearing in equation \eqref{eq:vectorized_TD_IO} reduce to identity matrices \cite{Rou_SPM_2024}, $i.e.,$ $\phi_\mathrm{CP}(n) = 0$ in equation \eqref{eq:diagonal_CP_matrix_def}, since there is no phase offset.

\vspace{-2ex}
\subsection{OTFS Signaling}

When \ac{OTFS} is used, multiple matrices $\mathbf{X}_u\in \mathcal{C}^{\tilde{K}\times \tilde{K}'}$ with $u = \{ 1,\dots,d_s \}$, containing a total of $\tilde{K} \tilde{K}' d_s$ symbols taken from an arbitrary complex constellation $\mathcal{C}$, are modulated as\footnote{For simplicity, we assume that all pulse-shaping operations utilize rectangular waveforms such that the corresponding sample matrices can be reduced to identity matrices.}
\vspace{-0.5ex}
\begin{equation}
\label{eq:TD_transmit_matrix_vectorized}
\mathbf{s}^{(\text{OTFS})}_u \triangleq \text{vec}\big(\mathbf{S}_u\big) = (\mathbf{F}_{\tilde{K}'}\herm \otimes \mathbf{I}_{\tilde{K}})  \text{vec}\big( \mathbf{X}_u
\big) \in \mathbb{C}^{\tilde{K}\tilde{K}'\times 1},
\end{equation}
where $\text{vec}(\cdot)$ denotes matrix vectorization via column stacking and $\mathbf{S}_u$ is a \ac{TD} symbols' matrix obtained from\footnote{Equivalently, $\mathbf{S}_u$ can be obtained as the Heisenberg transform of the \ac{ISFFT} of $\mathbf{X}_u$, $i.e.$, $\mathbf{S}_u = \mathbf{F}_{\tilde{K}}\herm \mathbf{X}_\text{FT}^u$ with $\mathbf{X}_\text{FT}^u \triangleq \mathbf{F}_{\tilde{K}} \mathbf{X}_u \mathbf{F}_{\tilde{K}'}\herm \in \mathbb{C}^{\tilde{K}\times \tilde{K}'}$.} the \ac{IDZT} of $\mathbf{X}_u$ as \cite{Hadani_WCNC_2017}
\begin{equation}
\label{eq:TD_transmit_matrix}
\mathbf{S}_u = \mathbf{X}_u \mathbf{F}_{\tilde{K}'}\herm  \in \mathbb{C}^{\tilde{K}\times \tilde{K}'}.
\end{equation}

We highlight that the notation in equation \eqref{eq:TD_transmit_matrix_vectorized} is in line with the strategy described in \cite{Raviteja_TWC_2018}, whereby the \ac{OTFS} signals are first vectorized and then appended with a \ac{CP} of length $N_\mathrm{CP}$ in order to eliminate inter-frame interference, in similarity with \ac{OFDM}. 
Taking advantage of this similarity, and in order to allow for direct comparisons between the two waveforms, we shall hereafter set $\tilde{K}\times \tilde{K}' = N$.

After transmission over the \ac{DD} channel $\bar{\mathbf{H}}(\bm{\mathcal{Z}},\tilde{\bm{\mathcal{Z}}},\bm{\mathcal{F}},\mathbf{V},\mathbf{U})$ as shown in equation \eqref{eq:vectorized_TD_IO_kron}, the $Nd_s$-element discrete-time received \ac{OTFS} signal can be modeled similar to equation \eqref{eq:TD_OFDM_input_output} as $\mathbf{r}_\text{OTFS} \triangleq \bar{\mathbf{H}}(\bm{\mathcal{Z}},\tilde{\bm{\mathcal{Z}}},\bm{\mathcal{F}},\mathbf{V},\mathbf{U})  \mathbf{s}_\text{OTFS} + \bar{\mathbf{w}}_\mathrm{TD}$, 
%
%
where the $Nd_s$-element vectors $\mathbf{s}_\text{OTFS}$ and $\mathbf{r}_\text{OTFS}$ are defined for \ac{OTFS} similar to equation \eqref{eq:OFDM_stacked_s}.
However, unlike \ac{OFDM}, the detection of the information symbols $\mathbf{X}_u$'s from the $\mathbf{r}^{(\text{OTFS})}_v$ elements $\forall$$v=1,\ldots,d_s$ of $\mathbf{r}_\text{OTFS}$ requires reversing the vectorization and the \ac{IDZT} operations employed in the construction of the $d_s$ elements of $\mathbf{s}_\text{OTFS}$, resulting in a distinct effective channel. 
In particular, let $\bm{R}_v \triangleq \text{vec}^{-1}(\mathbf{r}^{(\text{OTFS})}_v) \in \mathbb{C}^{\tilde{K} \times \tilde{K}'}$, with $\text{vec}^{-1}(\cdot)$ indicating the de-vectorization operation {whereby} a vector of size $\tilde{K}\tilde{K}' \times 1$ is reshaped into a matrix of size $\tilde{K} \times \tilde{K}'$, and consider the following \ac{DZT}\footnote{Equivalently, $\mathbf{Y}_v$ can be obtained as the SFFT of the Wigner transform of $\bm{R}_v$: $\mathbf{Y}_\text{FT}^v \triangleq \mathbf{F}_{\tilde{K}} \bm{R}_v$, yielding $\mathbf{Y}_v = \mathbf{F}_{\tilde{K}}\herm \mathbf{Y}_\text{FT}^v \mathbf{F}_{\tilde{K}'}\in \mathbb{C}^{\tilde{K} \times \tilde{K}'}$.}
\begin{equation}
\label{eq:DD_rec_sig_after_SFFT}
\mathbf{Y}_v  =  \bm{R}_v \mathbf{F}_{\tilde{K}'} \in \mathbb{C}^{\tilde{K} \times \tilde{K}'}.
\end{equation}

The demodulated \ac{OTFS} signal at the \ac{RX} then becomes
\begin{equation}
\label{eq:DD_demodulation}
\mathbf{y}^{(\text{OTFS})}_v \triangleq \text{vec}(\mathbf{Y}_v) = (\mathbf{F}_{\tilde{K}'} \otimes \mathbf{I}_{\tilde{K}})  \mathbf{r}^{(\text{OTFS})}_v \in \mathbb{C}^{N\times 1},
\end{equation}
which can be compactly written, similar to equation \eqref{eq:OFDM_input_output}, as the following $Nd_s$-element discrete-time received signal
\begin{equation}
\label{eq:DD_input_output_relation}
\mathbf{y}_\text{OTFS} = \bar{\mathbf{H}}_\text{OTFS}(\bm{\mathcal{Z}},\tilde{\bm{\mathcal{Z}}},\bm{\mathcal{F}},\mathbf{V},\mathbf{U})  \mathbf{x} + \bar{\mathbf{w}}_\text{OTFS},
\end{equation}
where $\bar{\mathbf{w}}_\text{OTFS} \in \mathbb{C}^{Nd_s \times 1}$ is an equivalent \ac{AWGN} with the same statistics as $\bar{\mathbf{w}}_\mathrm{TD}$, while $\bar{\mathbf{H}}_\text{OTFS}(\bm{\mathcal{Z}},\tilde{\bm{\mathcal{Z}}},\bm{\mathcal{F}},\mathbf{V},\mathbf{U}) \in \mathbb{C}^{Nd_s \times Nd_s}$ represents the effective \ac{OTFS} channel and is given by
\newpage

\quad\\[-6ex]
\begin{eqnarray}
\label{eq:OTFS_effective_channel}
\bar{\mathbf{H}}_\text{OTFS} \triangleq \sum_{p=1}^P \check{\mathbf{H}}_p^d \otimes \overbrace{( (\mathbf{F}_{\tilde{K}'} \otimes \mathbf{I}_{\tilde{K}}) \mathbf{G}_p  (\mathbf{F}_{\tilde{K}'}\herm \otimes \mathbf{I}_{\tilde{K}}))}^{\triangleq\mathbf{G}_p^\text{OTFS} \in \mathbb{C}^{N \times N}}&& \\[-1ex]
&&\hspace{-48ex}+ \sum_{k=1}^{K}  \sum_{\bar{p}=1}^{\bar{P}} \sum_{\tilde{p}=1}^{\tilde{P}} \check{\mathbf{H}}_{k,\bar{p},\tilde{p}}^\mathrm{RIS} \otimes \overbrace{( (\mathbf{F}_{\tilde{K}'} \otimes \mathbf{I}_{\tilde{K}}) \mathbf{G}_{k,\bar{p},\tilde{p}} (\mathbf{F}_{\tilde{K}'}\herm \otimes \mathbf{I}_{\tilde{K}}))}^{\triangleq\mathbf{G}_{k,\bar{p},\tilde{p}}^\text{OTFS} \in \mathbb{C}^{N \times N}} \nonumber \\
&&\hspace{-43ex}= \sum_{p=1}^P \check{\mathbf{H}}_p^d \otimes {\mathbf{G}_p^\text{OTFS}} + \sum_{k=1}^{K}  \sum_{\bar{p}=1}^{\bar{P}} \sum_{\tilde{p}=1}^{\tilde{P}} \check{\mathbf{H}}_{k,\bar{p},\tilde{p}}^\mathrm{RIS} \otimes {\mathbf{G}_{k,\bar{p},\tilde{p}}^\text{OTFS}}.\nonumber
\end{eqnarray}

Notice that similarly to the \ac{OFDM} case, the \ac{CP} phase matrices $\mathbf{\Theta}_p$'s reduce to identity matrices \cite{Rou_SPM_2024}.
Comparing the expressions in equation \eqref{eq:OFDM_effective_channel} and equation \eqref{eq:OTFS_effective_channel}, one can appreciate how \cite{Rou_SPM_2024}'s channel modeling approach elucidates both the similarity in form as well as the distinction in effect between the \ac{OFDM} and \ac{OTFS} waveforms in \ac{DD} channels.

\begin{figure*}[t!]
\centering
\captionsetup[subfloat]{labelfont=small,textfont=small}
\subfloat[\label{fig:TX-SIM_to_RX-SIM}
Integer Doppler frequencies $(f_1, f_2, f_3) = (0, -2, 1)$.]{
\begin{minipage}[b]{.32\textwidth}
\includegraphics[width=\textwidth]{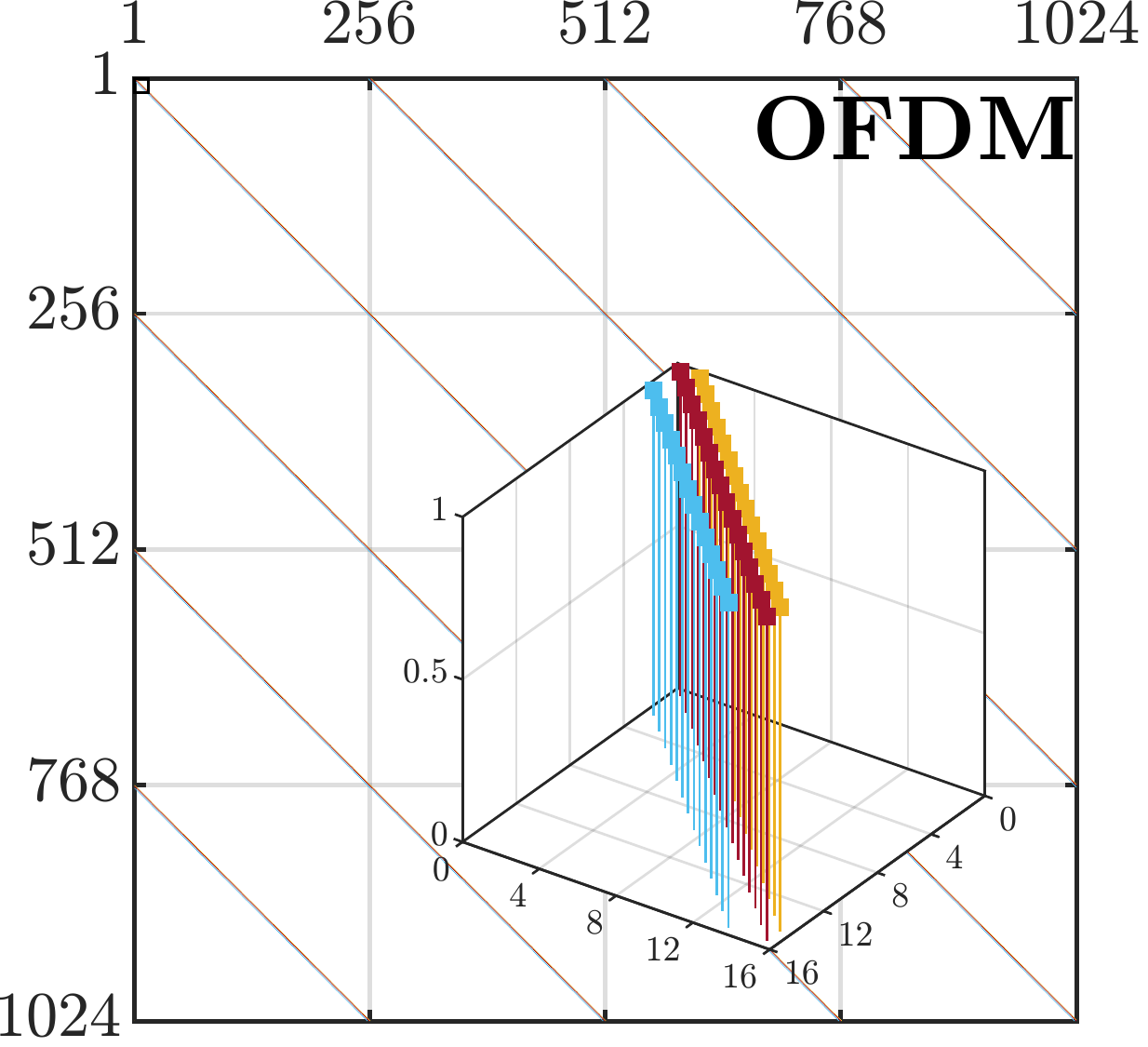}
\end{minipage}
\begin{minipage}[b]{.32\textwidth}
\includegraphics[width=\textwidth]{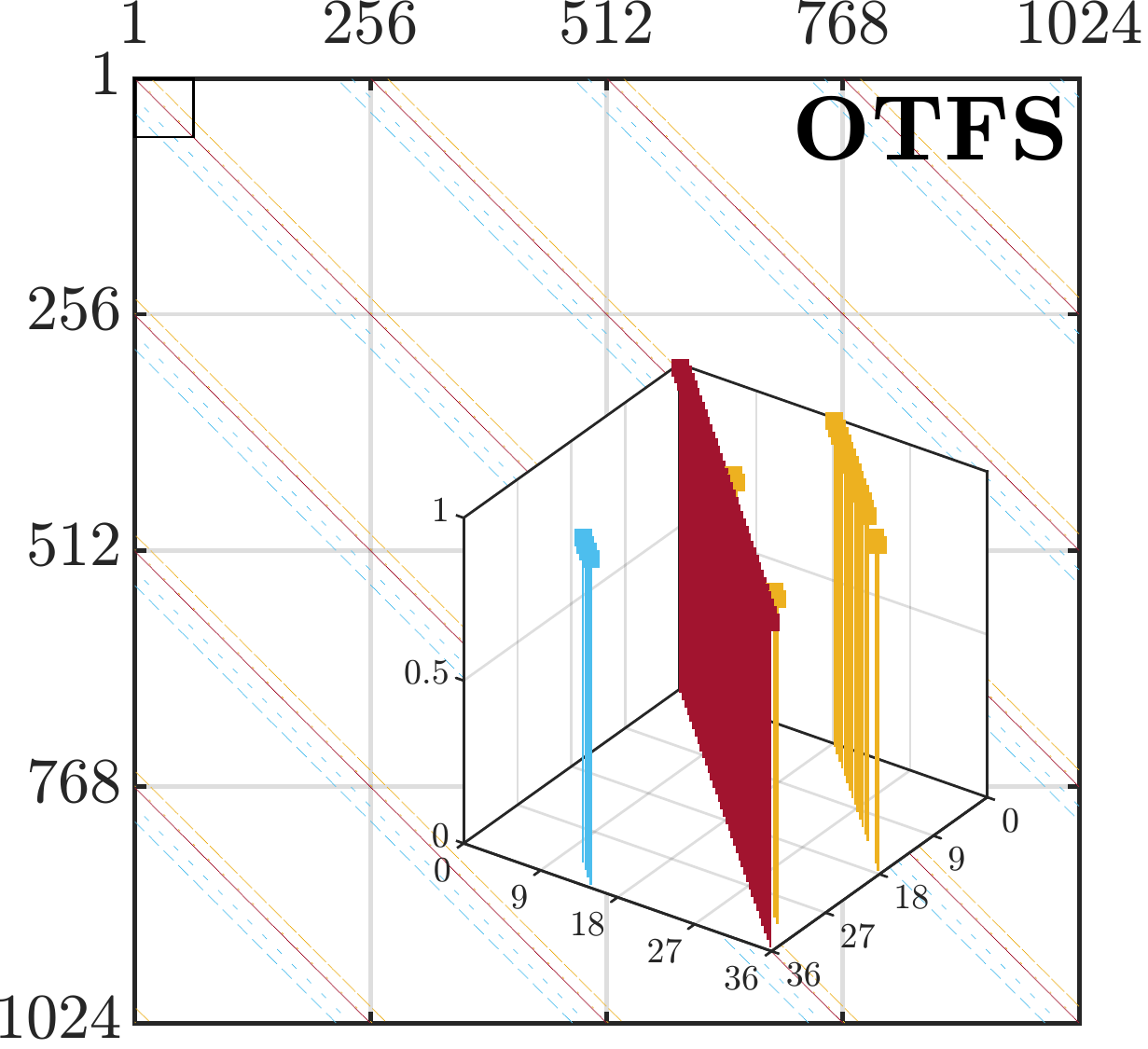}
\end{minipage}
\begin{minipage}[b]{.32\textwidth}
\includegraphics[width=\textwidth]{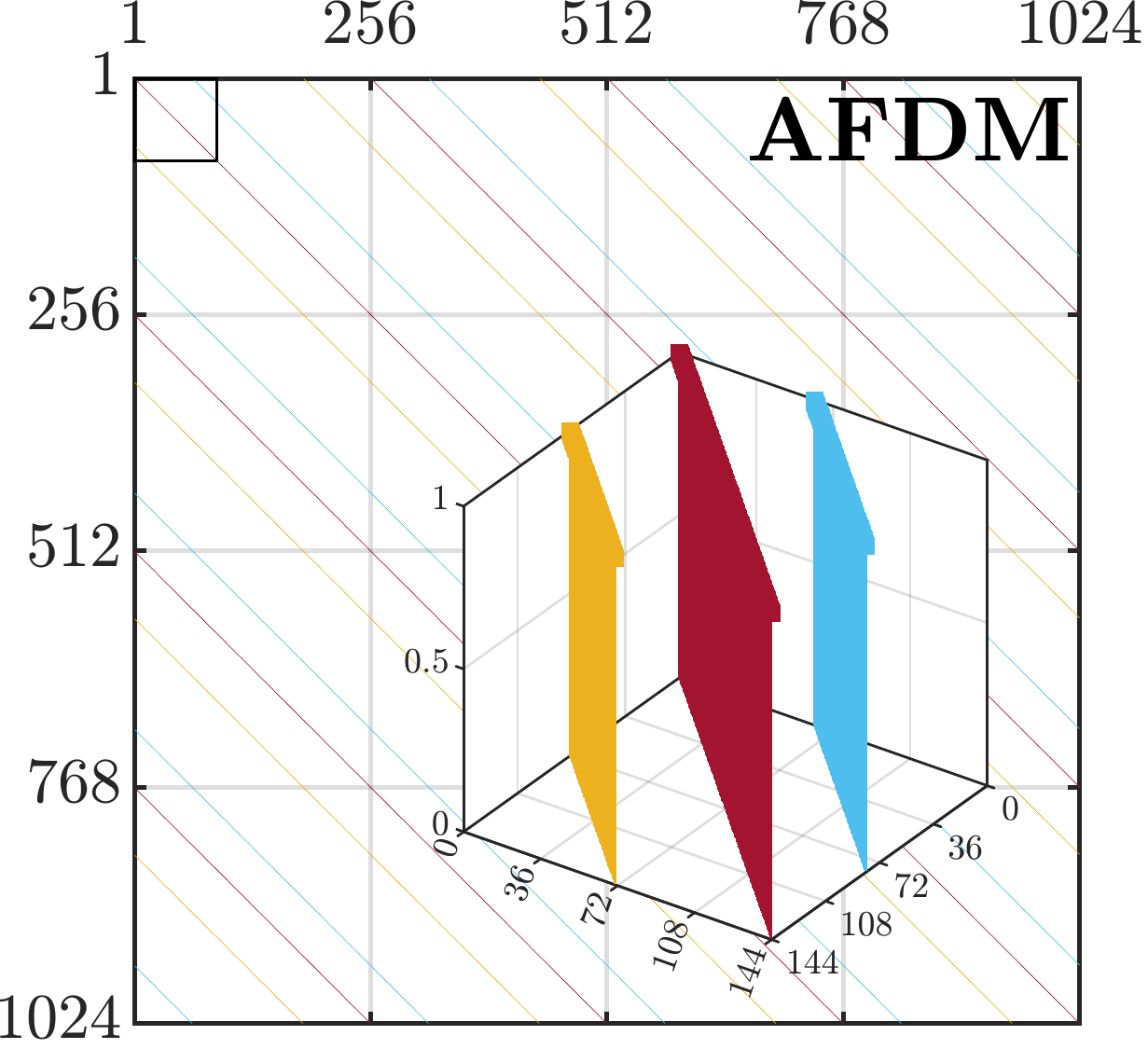}
\end{minipage}
}%
\par\bigskip
\vspace{-1ex}
\subfloat[\label{fig:TX-SIM_to_RX-SIM_frac}
Fractional Doppler frequencies $(f_1, f_2, f_3) = (0.698, -1.477, 1.124)$.]{
\begin{minipage}[b]{.32\textwidth}
\includegraphics[width=\textwidth]{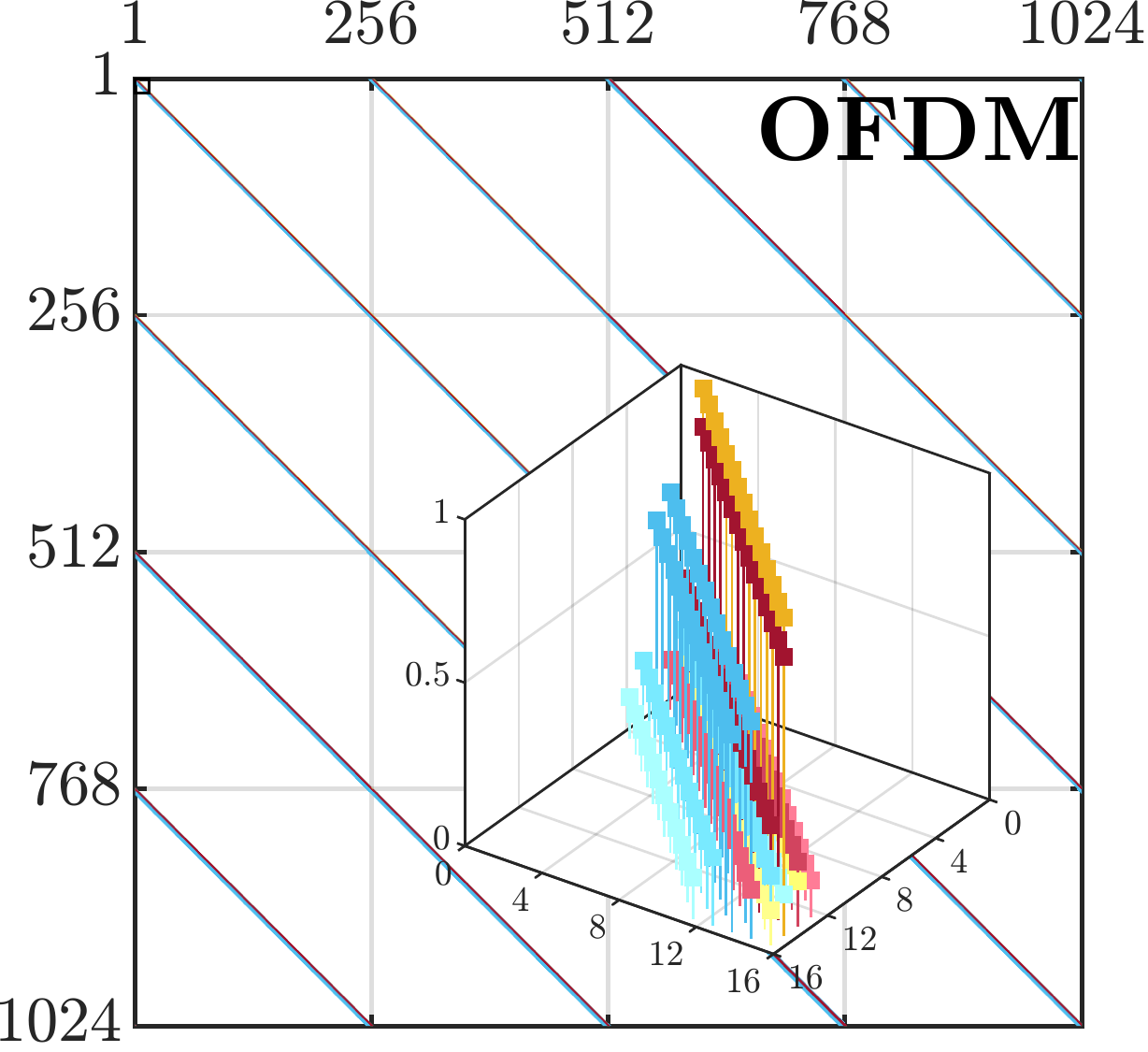}
\end{minipage}
\begin{minipage}[b]{.32\textwidth}
\includegraphics[width=\textwidth]{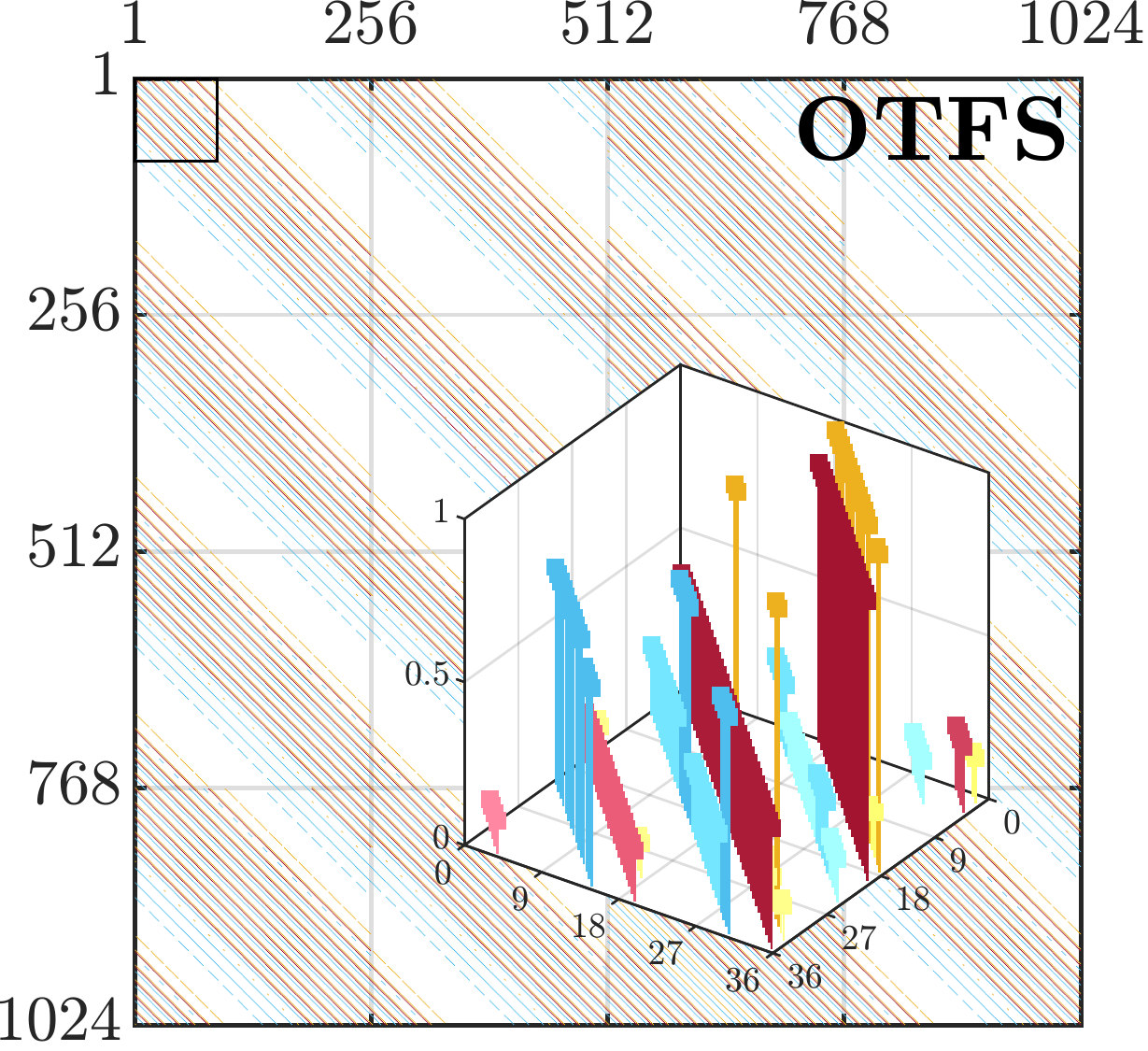}
\end{minipage}
\begin{minipage}[b]{.32\textwidth}
\includegraphics[width=\textwidth]{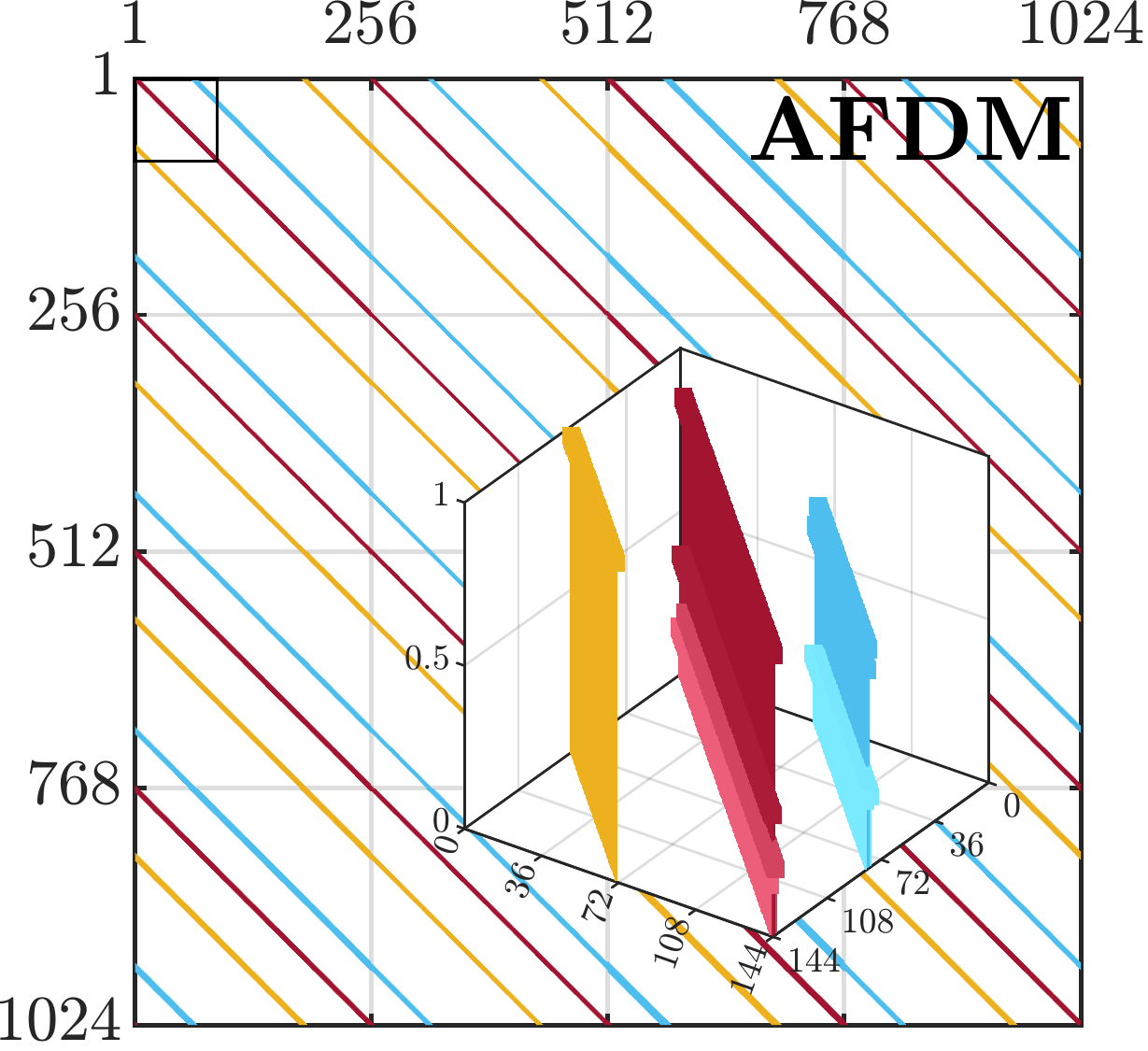}
\end{minipage}  
}%
\caption{Unoptimized $4\times4$ \ac{MPDD}-\ac{MIMO} with identical \ac{TX} and \ac{RX} \acp{SIM} with  $Q = \tilde{Q} = 5$ layers and $M = \tilde{M} = 100$ meta-atoms per layer, considering \ac{OFDM}, \ac{OTFS}, and \ac{AFDM} with $N=256$ symbols per frame and $P = 3$ channel paths with respective delays $(\ell_1, \ell_2, \ell_3) = (0, 5, 14)$ and integer (figure a) and fractional (figure b) Doppler frequencies.
{The x-axis and y-axis of each subfigure represents the row and column indices of the doubly-dispersive effective channel matrix $\bar{\mathbf{H}}$ for the various waveforms as defined in equations \eqref{eq:OFDM_effective_channel}, \eqref{eq:OTFS_effective_channel} and \eqref{eq:AFDM_effective_channel}.}
{Subsequently, t}he 3D inlays show the amplitude of the channel taps corresponding to the carriers outlined by the squares of the upper left corners.}
\label{fig:TX-SIM_to_RX-SIM_Full}
\vspace{-3ex}
\end{figure*}

\vspace{-1ex}
\subsection{AFDM Signaling}

The signal for transmission per information vector $\mathbf{x}_u$ when \ac{AFDM} waveform is used for the considered \ac{DD} \ac{MIMO} channel is given by  the \ac{IDAFT} as
\begin{equation}
\label{eq:AFDM_moduation}
\mathbf{s}^{(\text{AFDM})}_u \triangleq \mathbf{\Lambda}_1\herm  \mathbf{F}_{N}\herm  \mathbf{\Lambda}_2\herm  \mathbf{x}_u \in \mathbb{C}^{N \times 1},
\end{equation}
where the $N\times N$ matrices $\mathbf{\Lambda}_i$ with $i=1,2$ are defined as
\begin{equation}
\label{eq:lambda_def}
\mathbf{\Lambda}_i \triangleq \text{diag}\big(\big[1, e^{-\jmath2\pi c_i 2^2}, \ldots, e^{-\jmath2\pi c_i (N-1)^2}\big]\big),
\end{equation}
%
%
where the first central chirp frequency $c_1$ is an optimally designed parameter based on the maximum Doppler channel statistics \cite{Bemani_TWC_2023,Rou_SPM_2024}, while the second central chirp frequency $c_2$ is relatively a free parameter that can be exploited for \ac{ISAC} waveform shaping \cite{Zhu_Arxiv23} or information encoding \cite{Liu_Arxiv24,RouAsilimoar2024}.
%

It was shown in \cite{Rou_SPM_2024} that, after going through a \ac{DD} channel, an \ac{AFDM} modulated symbol vector $\mathbf{s}^{(\text{AFDM})}_u$ with the inclusion of a \ac{CPP} can be modeled similar to equation \eqref{eq:vectorized_TD_IO} by replacing the \ac{CP} matrix $\mathbf{\Theta}_p$ in equation \eqref{eq:diagonal_CP_matrix_def} with the \ac{CPP} matrix $\bm{\varTheta}_p$ given by equation \eqref{eq:AFDM_diagonal_CP_matrix_def} (top of {this} page). 
This implies that function $\phi_\mathrm{CP}(n)$ in equation \eqref{eq:diagonal_CP_matrix_def} needs to be set as $\phi_\mathrm{CP}(n) = c_1 (N^2 - 2Nn)$. 
To this end, the $Nd_s$-element discrete-time received \ac{AFDM} signal can be modeled similar to equation \eqref{eq:TD_OFDM_input_output} as $\mathbf{r}_\text{AFDM} \triangleq \bar{\mathbf{H}}(\bm{\mathcal{Z}},\tilde{\bm{\mathcal{Z}}},\bm{\mathcal{F}},\mathbf{V},\mathbf{U})  \mathbf{s}_\text{AFDM} + \bar{\mathbf{w}}_\mathrm{TD}$, where the $Nd_s$-element vectors $\mathbf{s}_\text{AFDM}$ and $\mathbf{r}_\text{AFDM}$ are defined for \ac{AFDM} similar to equation \eqref{eq:OFDM_stacked_s}. 
%
%
%
%
%
The \ac{AFDM} demodulation of each {element} $\mathbf{r}^{(\text{AFDM})}_v$ of $\mathbf{r}_\text{AFDM}${, with $v\in\{1,\ldots,d_s\}$,} is obtained as
\setcounter{equation}{39}
\begin{equation}
\mathbf{y}^{(\text{AFDM})}_v = \mathbf{\Lambda}_2  \mathbf{F}_{N}  \mathbf{\Lambda}_1  \mathbf{r}^{(\text{AFDM})}_v \in \mathbb{C}^{N\times 1},
\label{eq:AFDM_demodulation}
\end{equation}
yielding the following expression for the $Nd_s{\times1}$ discrete-time received signal{, similar to equations \eqref{eq:OFDM_input_output} and \eqref{eq:DD_input_output_relation}}
\begin{equation}
\mathbf{y}_\text{AFDM} = \bar{\mathbf{H}}_\text{AFDM}(\bm{\mathcal{Z}},\tilde{\bm{\mathcal{Z}}},\bm{\mathcal{F}},\mathbf{V},\mathbf{U})  \mathbf{x} + \bar{\mathbf{w}}_\mathrm{AFDM},
\label{eq:DAF_input_output_relation}
\end{equation}
where $\bar{\mathbf{w}}_\mathrm{AFDM} \in \mathbb{C}^{Nd_s \times 1}$ is an equivalent \ac{AWGN} holding the same statistics with $\bar{\mathbf{w}}_\mathrm{TD}$, and $\bar{\mathbf{H}}_\text{AFDM}(\bm{\mathcal{Z}},\tilde{\bm{\mathcal{Z}}},\bm{\mathcal{F}},\mathbf{V},\mathbf{U}) \in \mathbb{C}^{Nd_s \times Nd_s}$ indicates the effective \ac{AFDM} channel given by
\begin{eqnarray}
\label{eq:AFDM_effective_channel}
\bar{\mathbf{H}}_\text{AFDM} \triangleq \sum_{p=1}^P \check{\mathbf{H}}_p^d \otimes \overbrace{( \mathbf{\Lambda}_2  \mathbf{F}_{N}  \mathbf{\Lambda}_1 \mathbf{G}_p \mathbf{\Lambda}_1\herm  \mathbf{F}_{N}\herm  \mathbf{\Lambda}_2\herm)}^{\mathbf{G}_p^\text{AFDM} \in \mathbb{C}^{N \times N}}&& \\[-1.5ex]
&&\hspace{-42ex}+ \sum_{k=1}^{K}  \sum_{\bar{p}=1}^{\bar{P}} \sum_{\tilde{p}=1}^{\tilde{P}} \check{\mathbf{H}}_{k,\bar{p},\tilde{p}}^\mathrm{RIS} \otimes \overbrace{( \mathbf{\Lambda}_2  \mathbf{F}_{N}  \mathbf{\Lambda}_1 \mathbf{G}_{k,\bar{p},\tilde{p}} \mathbf{\Lambda}_1\herm  \mathbf{F}_{N}\herm  \mathbf{\Lambda}_2\herm)}^{\mathbf{G}_{k,\bar{p},\tilde{p}}^\text{AFDM} \in \mathbb{C}^{N \times N}} \nonumber \\[-0.25ex]
&&\hspace{-40ex}=  \sum_{p=1}^P \check{\mathbf{H}}_p^d \otimes {\mathbf{G}_p^\text{AFDM}} + \sum_{k=1}^{K}  \sum_{\bar{p}=1}^{\bar{P}} \sum_{\tilde{p}=1}^{\tilde{P}} \check{\mathbf{H}}_{k,\bar{p},\tilde{p}}^\mathrm{RIS} \otimes {\mathbf{G}_{k,\bar{p},\tilde{p}}^\text{AFDM}}.\nonumber 
\end{eqnarray}
%

Clearly, equation \eqref{eq:AFDM_effective_channel} has the same structure of equations \eqref{eq:OFDM_effective_channel} and \eqref{eq:OTFS_effective_channel}, with the same holding for the \ac{MIMO} input-output relationships described by equations \eqref{eq:OFDM_input_output}, \eqref{eq:DD_input_output_relation} and \eqref{eq:DAF_input_output_relation}. 
This implies that signal processing techniques such as channel estimation can be designed to under a unified framework, applying to \ac{OFDM}, \ac{OTFS}, \ac{AFDM}, and similar waveforms.

Finally, for the sake of clarity, we emphasize that a ``conventional'' \ac{DD}-\ac{MIMO} model -- $i.e.$, a \ac{MIMO} extension of the model in \cite{Rou_SPM_2024} without the incorporation of \ac{RIS} in the ambient and of \ac{TX} and \ac{RX} \acp{SIM} -- can be trivially extracted from the above.
For instance, for the \ac{OFDM}, \ac{OTFS}, and \ac{AFDM} waveforms, equations \eqref{eq:OFDM_effective_channel}, \eqref{eq:OTFS_effective_channel}, and \eqref{eq:AFDM_effective_channel}, would yield
\begin{equation}
\label{eq:H_DD_MIMO}
\!\bar{\mathbf{H}}_\text{MIMO}\!\triangleq \!\sqrt{\tfrac{N_\mathrm{T} N_\mathrm{R}}{P}}\sum_{p=1}^P\! \left(h_p     \mathbf{a}_\mathrm{R}\left(\phi_p^{\rm in}\right)\! \mathbf{a}_\mathrm{T}\herm\left(\phi_p^{\rm out}\right)\right) \otimes {\mathbf{G}_p^\text{MIMO}},\!
\end{equation}
where the previous subscripts \ac{OFDM}, \ac{OTFS}, and \ac{AFDM}, are respectively represented by the generic subscript \ac{MIMO}.

%
%

\vspace{-2ex}
\subsection{Comparison of Effective Channels}
\label{subsec:comp_eff_channel}

As an example of how the aforementioned waveforms are affected by a MPDD \ac{MIMO} channel, consider a point-to-point system with $N_\mathrm{T} = N_\mathrm{R} = 4$ antennas, such that up to $d_s = 4$ independent data streams can be utilized.
Assume that the system operates with a carrier frequency of $28$ GHz ($i.e.$, at wavelength $\lambda = 10.7$ mm) with a bandwidth $B = 20$ MHz in an environment with identical \ac{TX} and \ac{RX} \acp{SIM}, both with $Q = \tilde{Q} = 5$ layers of metasurfaces and $M = \tilde{M} = 100$ meta-atoms per layer, leading to $M_x = M_z = \tilde{M}_x = \tilde{M}_z = 10$. 
In line with the Rayleigh-Sommerfeld diffraction theory, we assume that the distance between any two adjacent metasurface layers at either the \ac{TX} or \ac{RX} is $5\lambda$, while the distance between two adjacent meta-atoms on any layer is $\lambda/2$ both along the $x$- and $z$-axis, such that $\rho_t = \rho_r = \lambda^2/4$. 
We assume that the \acp{SIM} are unoptimized and that the environment lacks any \ac{RIS} presence, thus, we set $\bm{\mathcal{Z}} = \tilde{\bm{\mathcal{Z}}} = \bm{\mathcal{F}} = \mathbf{I}_{M/\tilde{M}/J}, \forall q, \tilde{q}, k$.

Since the \acp{ULA} are aligned with the meta-atoms on the \acp{SIM}, the angle between the propagation and the normal direction of the metasurface layers becomes in this case $\epsilon^{q}_{m,m'} = \tilde{\epsilon}^{\tilde{q}}_{\tilde{m},\tilde{m}'} = 0$ $\forall q,\tilde{q},m,\tilde{m},m',\tilde{m}'$.
Consequently, $\bm{\Psi}_q$'s, $\bm{\Gamma}_q$'s, $\bm{\Delta}_q$'s, and $\bm{\Xi}_q$'s can be generated from equations \eqref{eq:diagona_shift_matrix_per_layer} and \eqref{eq:diffraction_coeff}, leading to the generation of the transfer functions $\bm{\Upsilon}_\mathrm{T}$ and $\bm{\Upsilon}_\mathrm{R}$ for the \ac{TX} and \ac{RX} \ac{SIM}, respectively, as given in equations \eqref{eq:transmit_SIM_full} and \eqref{eq:receive_SIM_full}, respectively. 

The \ac{TX}/\ac{RX} digital beamformers are also unoptimized and set as $\mathbf{V} = \mathbf{U} = \mathbf{I}_{d_s}$. 
Finally, the sampling frequency is set to $F_\mathrm{S}=B$ and the number of symbols per frame to $N = 256$.   

For the \ac{MPDD} channel, we assume that the path delays $\tau_p$'s are uniformly distributed in $[0,\tau_\text{max}]$, and that the Doppler shifts follow a Jakes spectrum, $i.e.$, $\nu_p = \nu_\text{max} \cos(\theta_p)$ $\forall p$ with each $\theta_p$ uniformly distributed in $[-\pi,\pi]$. 
In turn, we assume that the \ac{2D} and \ac{3D} elevation \acp{AoD}/\acp{AoA} are uniformly distributed in $[0,\pi]$, while the \ac{3D} azimuth \acp{AoD}/\acp{AoA} are in $[-\frac{\pi}{2},\frac{\pi}{2}]$. 
Finally, we consider a case with $P = 3$ paths, with respective delays $[\ell_1, \ell_2, \ell_3] = [0, 5, 14]$. 

Figure \ref{fig:TX-SIM_to_RX-SIM_Full} illustrates the resulting \underline{unoptimized} \ac{MPDD}-\ac{MIMO} channels with all considered waveforms. 
As observed from Fig. \ref{fig:TX-SIM_to_RX-SIM_frac}, the fractional components of the Doppler shift ``spread'' the path-wise components, making the effective channel matrix design vital for estimation/detection tasks to avoid overlaps and mixing between paths.
Comparing the results shown in Figure \ref{fig:TX-SIM_to_RX-SIM_Full} with those in \cite{Rou_SPM_2024} it can be confirmed that, as expected, unoptimized \acp{SIM} have no effect onto the \ac{DD} channels undergone by the compared \ac{OFDM}, \ac{OTFS} and \ac{AFDM} waveforms.
In what follows we will demonstrate, however, that when optimized, these structures can significantly impact on the detection performance of such systems.

\vspace{-1ex}
\section{\ac{SIM} Optimization and Data Detection}
\label{sec:joint_optimization}

A great potential advantage of \ac{MIMO} systems incorporating \acp{SIM}, compared to conventional \ac{MIMO} systems, is that signal processing functions previously carried out by circuitry, be it in analog or digital fashions, can instead be performed passively at the wave domain \cite{AnJSAC2023}.
And while this new wave-domain processing capability can be exploited to \underline{replace} classic digital/analog processing, as suggested $e.g.$ in \cite{AnWC2024}, it can also be utilized to \underline{augment} it.
Focusing on the latter case, and to elaborate further, consider for example the trade-off that exists between enhancing the \ac{SNR} of received signals and improving communication rate \cite{TseTIT2004}, which from the viewpoint of receiver design\footnotemark, translates to either designing directivity-enhancing receive beamformers ($i.e.$, combiners) or, instead, exploiting the multiple streams of data as extrinsic information for detection \cite{MolischCOMMMAG2017}.


Under the conventional paradigm, a choice (or trade-off) between these competing interests must be made.
In contrast, in the case of \ac{MIMO}-\ac{SIM} systems, one can seek to reap the best of both worlds, by parameterizing the \ac{SIM} for \ac{SNR} gain, while leaving the degrees of freedom afforded by the multiple \ac{RX} antennas to design robust detectors.

With the latter approach in mind, we offer in this section an illustrative application of the channel model detailed above.
In particular, we first formulate an optimization problem that leverages the reconfigurability of the proposed \ac{MPDD}-\ac{MIMO} model to increase the intensity of the complex channel coefficients at the \ac{EM} domain,
and subsequently introduce a GaBP-based data detection algorithm that exploits the signals from all \ac{RX} antennas simultaneously and extrinsically.
As a bonus, the flexibility of the model is highlighted by offering both of the aforementioned contributions in a manner that they apply to \ac{OFDM}, \ac{OTFS} and \ac{AFDM} alike. 

\footnotetext{It is well known that \ac{RX} \ac{BF} does not contribute to rate achievement in MIMO systems \cite{SandovalACCESS2023}.}

\vspace{-2ex}
\subsection{SIM-based Signal Enhancement}
\label{SIM_Optimization}

Referring to equation \eqref{eq:sampled_TD}, setting the \ac{BF} matrices $\bm{U}$ and $\bm{V}$ to identities to put emphasis on the impact of \acp{SIM} as per the discussion above, and considering for simplicity a scenario without \ac{RIS}, we seek to parametrize the matrices $\bm{\mathcal{Z}}$ and $\tilde{\bm{\mathcal{Z}}}$ to enhance the total receive signal power, which can be achieved by solving the optimization problem\footnote{Notice that, although leading to significantly improvement in sensing and communication performances in the \ac{DD} channel, the optimization problem here proposed is not impacted by \ac{DD} effects under the model described by in Section \ref{IO_Model}, which further validates the overall contribution of the article.}
\vspace{-1ex}
\begin{align}
\label{eq:full_optimization_problem_channel_coeff}
&\underset{\bm{\mathcal{Z}},\tilde{\bm{\mathcal{Z}}}}{\text{max}} \; \mathcal{O}(\bm{\mathcal{Z}},\tilde{\bm{\mathcal{Z}}}) = \sum_{p=1}^P \Big|\Big| \overbrace{\tilde{h}_p \bm{\Upsilon}_\mathrm{R}(\tilde{\bm{\mathcal{Z}}}) \mathbf{R}_\mathrm{RX}^{1/2} \mathbf{B}_p \mathbf{R}_\mathrm{TX}^{1/2} \bm{\Upsilon}_\mathrm{T}(\bm{\mathcal{Z}})}^{\triangleq\,\mathbf{O}_p}  \Big|\Big|_F^2\nonumber\\
&\;\;\;\text{\text{s}.\text{t}.}\;\bm{\Upsilon}_\mathrm{T}(\bm{\mathcal{Z}})\;\text{as in equation \eqref{eq:transmit_SIM_full}},\;\bm{\Upsilon}_\mathrm{R}(\tilde{\bm{\mathcal{Z}}})\;\text{as in equation \eqref{eq:receive_SIM_full}},\nonumber\\
&\;\;\;\quad\;\;\,\bm{\Psi}_q \; \text{and} \; \bm{\Delta}_{\tilde{q}}\;\text{as in eqs. \eqref{eq:diagona_shift_matrix_per_layer} and \eqref{eq:diagona_shift_matrix_per_layer_rx}, respectively},\nonumber\\
&\;\;\;\quad\;\;\,|\zeta^{q}_{m}| \leq \pi \; \forall (q,m), \text{and\;} \,|\tilde{\zeta}^{\tilde{q}}_{\tilde{m}}| \leq \pi \; \forall (\tilde{q},\tilde{m}).
\end{align}

Notice that in the above  we have used the array response matrix $\bm{B}_p$ corresponding to a $p$-th path, defined previously in equation \eqref{eq:Bmatrix}, as well as the scalar $\tilde{h}_p$ defined as
\vspace{-0.5ex}
\begin{equation}
\label{eq:h_tilde_p_and_p_k}
\tilde{h}_p \triangleq h_p  \sqrt{\tfrac{M \tilde{M}}{P}}.\,\,
\vspace{-0.5ex}
\end{equation}


In view of the non-convex unit modulus constraints in equation \eqref{eq:full_optimization_problem_channel_coeff}, we utilize a simple gradient ascent technique to tune the phase shift parameters of the \acp{SIM}.
In particular, following \cite{AnICC2024}, the gradient ascent algorithm is employed to adjust the phase shifts of the transmit and receive \acp{SIM} iteratively, maximizing the objective function in equation \eqref{eq:full_optimization_problem_channel_coeff}. 
The algorithm is divided into two main discrete steps, addressed in detailed below.

\subsubsection{Gradient Calculation} The gradient of the objective function $\mathcal{O}(\bm{\mathcal{Z}},\tilde{\bm{\mathcal{Z}}})$ with respect to the phase shift vector of the $q$-th layer of the \ac{TX}-\ac{SIM}, denoted by $\bm{\zeta}_{q} = [\zeta^{q}_{1}, \zeta^{q}_{2}, \dots, \zeta^{q}_{M}]\trans $, is given by
\vspace{-1ex}
\begin{equation}
\label{eq:gradient_def_tx}
\nabla_{\bm{\zeta}_{q}} \mathcal{O}(\bm{\mathcal{Z}},\tilde{\bm{\mathcal{Z}}}) = \sum_{p=1}^P \sum_{n_t=1}^{N_T} \nabla_{\bm{\zeta}_{q}} ||\bm{o}_{p,n_t}||^2, \; \forall q,
\end{equation}
where $\bm{o}_{p,n_t} \in \mathbb{C}^{N_R \times 1}$, with $n_t = \{1, 2, \dots, N_T\}$ and $p  = \{1, 2, \dots, P\}$, represents the $n_t$-th column of the $p$-th $\mathbf{O}_{p}$ matrix implicitly defined in the bracket of equation \eqref{eq:full_optimization_problem_channel_coeff}.

Similarly, the gradient with respect to the phase shift vector $\tilde{\bm{\zeta}}_{\tilde{q}} = [ \tilde{\zeta}^{\tilde{q}}_{1}, \tilde{\zeta}^{\tilde{q}}_{2}, \dots, \tilde{\zeta}^{\tilde{q}}_{\tilde{M}} ]\trans$ of the $\tilde{q}$-th layer of the \ac{RX}-\ac{SIM} is given by

\quad\\[-5ex]
\begin{equation}
\label{eq:gradient_def_rx}
\nabla_{\tilde{\bm{\zeta}}_{\tilde{q}}} \mathcal{O}(\bm{\mathcal{Z}},\tilde{\bm{\mathcal{Z}}}) = \sum_{p=1}^P \sum_{n_t=1}^{N_T} \nabla_{\tilde{\bm{\zeta}}_{\tilde{q}}} ||\bm{o}_{p,n_t}||^2, \; \forall \tilde{q}.
\end{equation}

Leveraging the chain rule, the per-shift partial derivates of $||\bm{o}_{p,n_t}||^2$ with respect to $\zeta^{q}_{m}$ are given as
\vspace{-1ex}
\begin{align}
\label{eq:partial_tx_o}
\frac{\partial ||\bm{o}_{p,n_t}||^2}{\partial \zeta^{q}_{m}} & = 2 \Re \bigg\{ \frac{\partial \bm{o}_{p,n_t} \herm }{ \partial \zeta^{q}_{m} }  \bm{o}_{p,n_t} \bigg\} \nonumber \\
& = 2 \Re \bigg\{ \frac{\partial \big( \tilde{\bm{\Upsilon}}_{t:q,p,n_t}  \vect{\bm{\Psi}_q} \big) \herm }{ \partial \zeta^{q}_{m} }  \bm{o}_{p,n_t} \bigg\} \nonumber \\
& = 2 \Re \Big\{ -\jmath e^{-\jmath \zeta^{q}_{m}} \mathbf{i}_{m}\trans \tilde{\bm{\Upsilon}}_{t:q,p,n_t}\herm  \bm{o}_{p,n_t} \Big\} \nonumber \\
& = 2 \Im \Big\{ e^{-\jmath \zeta^{q}_{m}} \mathbf{i}_{m}\trans \tilde{\bm{\Upsilon}}_{t:q,p,n_t}\herm  \bm{o}_{p,n_t} \Big\}, \forall m,q,
\end{align}
where $\mathbf{i}_m$ stands for the $m$-th column of $\mathbf{I}_{M}$ and the second equatlity holds due to $\bm{o}_{p,n_t} = \tilde{\bm{\Upsilon}}_{t:q,p,n_t} \vect{\mathbf{\Psi}_{q}}$, with $\tilde{\bm{\Upsilon}}_{t:q,p,n_t} \in \mathbb{C}^{N_R \times M}$ denoting the equivalent channel matrix of the $p$-th path associated to the $q$-th layer of the \ac{TX}-\ac{SIM} and the $n_t$-th transmit antenna, which is defined as
\vspace{-1ex}
\begin{align}
  \label{eq:Upsilon_t}
\tilde{\bm{\Upsilon}}_{t:q,p,n_t} \triangleq& \tilde{h}_p \bm{\Upsilon}_\mathrm{R}(\tilde{\bm{\mathcal{Z}}}) \mathbf{R}_\mathrm{RX}^{1/2} \mathbf{B}_p \mathbf{R}_\mathrm{TX}^{1/2} \prod_{q'=1}^{q+1} \bm{\Psi}_{Q-q'+1} \bm{\Gamma}_{Q-q'+1} \nonumber \\[-1ex]
&\times \text{diag}(\bm{s}_{q,n_t}),
\end{align}
where $\bm{s}_{q,n_t} \in \mathbb{C}^{M \times 1}$ is the signal component activating the $q$-th layer of the \ac{TX}-\ac{SIM} associated to the $n_t$-th transmit antenna, which is defined as the $n_t$-th column of 
\vspace{-1ex}
\begin{equation}
  \label{S_TX_part}
\bm{S}_{q} = \bm{\Gamma}_{q}\!\!\!\!\!\!\! \prod_{q'= Q - q + 2}^{Q}\!\!\!\!\!\!\! \bm{\Psi}_{Q-q'+1} \bm{\Gamma}_{Q-q'+1}.
\end{equation}

Finally, the $M$ partial derivatives of each $p$-th path can be gathered into a vector, yielding the following final expression for the gradient of the \ac{TX}-\ac{SIM}
\vspace{-1ex}
\begin{equation}
\label{eq:final_grad_TXSIM}
\nabla_{\bm{\zeta}_{q}} \mathcal{O}(\bm{\mathcal{Z}},\tilde{\bm{\mathcal{Z}}}) = 2 \Im \bigg\{ \sum_{p=1}^P \sum_{n_t=1}^{N_T} \bm{\Psi}_{q}\herm \tilde{\bm{\Upsilon}}_{t:q,p,n_t}\herm \bm{o}_{p,n_t} \bigg\}.
\end{equation}

Similarly, considering the definition of the gradient for the \ac{RX}-\ac{SIM} in equation \eqref{eq:gradient_def_rx} and expressing the per-shift partial derivates with respect to $\zeta^{\tilde{q}}_{\tilde{m}}$ yields
\begin{align}
\label{eq:partial_rx_o}
\frac{\partial ||\bm{o}_{p,n_t}||^2}{\partial \zeta^{\tilde{q}}_{\tilde{m}}} & = 2 \Re \bigg\{ \frac{\partial \bm{o}_{p,n_t} \herm }{ \partial \zeta^{\tilde{q}}_{\tilde{m}} }  \bm{o}_{p,n_t} \bigg\} \nonumber \\
& = 2 \Re \bigg\{ \frac{\partial \big( \tilde{\bm{\Upsilon}}_{r:\tilde{q},p,n_t}  \vect{\tilde{\bm{\Delta}}_{\tilde{q}}} \big) \herm }{ \partial \zeta^{\tilde{q}}_{\tilde{m}} }  \bm{o}_{p,n_t} \bigg\} \nonumber \\
& = 2 \Re \Big\{ -\jmath e^{-\jmath \zeta^{\tilde{q}}_{\tilde{m}}} \mathbf{i}_{\tilde{m}}\trans \tilde{\bm{\Upsilon}}_{r:\tilde{q},p,n_t}\herm  \bm{o}_{p,n_t} \Big\} \nonumber \\
& = 2 \Im \Big\{ e^{-\jmath \zeta^{\tilde{q}}_{\tilde{m}}} \mathbf{i}_{\tilde{m}}\trans \tilde{\bm{\Upsilon}}_{r:\tilde{q},p,n_t}\herm  \bm{o}_{p,n_t} \Big\}, \forall \tilde{m},\tilde{q},
\end{align}
where $\mathbf{i}_{\tilde{m}}$ stands for the $\tilde{m}$-th column of $\mathbf{I}_{\tilde{M}}$ and the second equality holds due to $\bm{o}_{p,n_t} = \tilde{\bm{\Upsilon}}_{r:\tilde{q},p,n_t} \vect{\mathbf{\Delta}_{\tilde{q}}}$ with $\tilde{\bm{\Upsilon}}_{r:\tilde{q},p,n_t} \in \mathbb{C}^{N_R \times \tilde{M}}$ denoting the equivalent channel matrix of the $p$-th path associated to the $\tilde{q}$-th layer of the RX-\ac{SIM} and the $n_t$-th transmit antenna, which is defined as
\vspace{-1ex}
\begin{equation}
  \label{eq:Upsilon_r}
\tilde{\bm{\Upsilon}}_{r:\tilde{q},p,n_t} \triangleq \tilde{h}_p \bm{\Xi}_{1} \bigg( \prod_{\tilde{q}'=1}^{\tilde{q} - 1} \bm{\Delta}_{\tilde{q}'} \bm{\Xi}_{\tilde{q}' + 1} \bigg) \text{diag}(\tilde{\bm{s}}_{\tilde{q},n_t}) ,
\end{equation}
where $\tilde{\bm{s}}_{\tilde{q},n_t} \in \mathbb{C}^{\tilde{M} \times 1}$ is the signal component activating the $\tilde{q}$-th layer of the \ac{RX}-\ac{SIM} associated to the $n_t$-th transmit antenna, which is defined as the $n_t$-th column of the matrix
\begin{equation}
  \label{S_RX_part}
\tilde{\bm{S}}_{\tilde{q}} = \bigg( \prod_{\tilde{q}'= \tilde{q} + 1}^{\tilde{Q}} \bm{\Xi}_{\tilde{q}'} \bm{\Delta}_{\tilde{q}'} \bigg) \mathbf{R}_\mathrm{RX}^{1/2} \mathbf{B}_p \mathbf{R}_\mathrm{TX}^{1/2} \bm{\Upsilon}_\mathrm{T}(\bm{\mathcal{Z}}).
\end{equation}

Finally, the $\tilde{M}$ partial derivatives of each $p$-th path can be gathered, giving the final calculation for the gradient of the \ac{TX}-\ac{SIM} as
\begin{equation}
\label{eq:final_grad_RXSIM}
\nabla_{\tilde{\bm{\zeta}}_{\tilde{q}}} \mathcal{O}(\bm{\mathcal{Z}},\tilde{\bm{\mathcal{Z}}}) = 2 \Im \bigg\{ \sum_{p=1}^P \sum_{n_t=1}^{N_T} \bm{\Delta}_{\tilde{q}}\herm \tilde{\bm{\Upsilon}}_{r:\tilde{q},p,n_t}\herm \bm{o}_{p,n_t} \bigg\}.
\end{equation}

\subsubsection{Parameter Update} With the aforementioned closed-form expressions for the gradient of $\mathcal{O}(\bm{\mathcal{Z}},\tilde{\bm{\mathcal{Z}}})$ with respect to the \ac{TX}- and \ac{RX}-\acp{SIM} phase parameters respectively given by equations \eqref{eq:final_grad_TXSIM} and \eqref{eq:final_grad_RXSIM},
the update required to iteratively adjust the phases $\bm{\zeta}_{q}$ and $\tilde{\bm{\zeta}}_{\tilde{q}}$ to optimize total receive power, as described in equation \eqref{eq:full_optimization_problem_channel_coeff}, can be efficiently computed by 
\begin{subequations}
\label{eq:param_updates_GD}
\begin{equation}
\bm{\zeta}_{q}^{(i + 1)} = \bm{\zeta}_{q}^{(i)} + \lambda^{(i)} \rho^{(i)} \nabla_{\bm{\zeta}_{q}} \mathcal{O}(\bm{\mathcal{Z}},\tilde{\bm{\mathcal{Z}}}),
\end{equation}
\begin{equation}
\tilde{\bm{\zeta}}_{\tilde{q}}^{(i + 1)} = \tilde{\bm{\zeta}}_{\tilde{q}}^{(i)} + \lambda^{(i)} \tilde{\rho}^{(i)} \nabla_{\tilde{\bm{\zeta}}_{\tilde{q}}} \mathcal{O}(\bm{\mathcal{Z}},\tilde{\bm{\mathcal{Z}}}),
\end{equation}
\end{subequations}
where $\lambda^{(i)} \in (0,1)$ is the decaying learning rate parameter to ensure convergence and $\rho^{(i)},\tilde{\rho}^{(i)}$ are normalization parameters calculated at each step as
\begin{subequations}
\label{eq:normalization_params}
\begin{equation}
\rho^{(i)} = \pi / \underset{q \in Q, m \in M}{\max}\nabla_{\bm{\zeta}_{q}} \mathcal{O}(\bm{\mathcal{Z}},\tilde{\bm{\mathcal{Z}}}),
\end{equation}
\begin{equation}
\tilde{\rho}^{(i)} = \pi / \underset{\tilde{q} \in \tilde{Q}, \tilde{m} \in \tilde{M}}{\max}\nabla_{\tilde{\bm{\zeta}}_{\tilde{q}}} \mathcal{O}(\bm{\mathcal{Z}},\tilde{\bm{\mathcal{Z}}}).
\end{equation}
\end{subequations}

\vspace{-2ex}
\subsection{GaBP-based Data Detection}
\label{subsec:GaBP}

In possession of the \ac{SIM} optimization method detailed above, and given the  input-output relationships given in Section \ref{IO_Model} for various exemplary \ac{ISAC}-enabling waveforms, we finally seek to illustrate the impact of integrating \acp{SIM} onto the design of communication systems under the \ac{MPDD}-\ac{MIMO} channel model described in Section \ref{MPDD_MIMO_Model}, by comparing the corresponding performances of \ac{OFDM}, \ac{OTFS} and \ac{AFDM}, with and without \acp{SIM}.

Before we proceed, we recall that it has been widely demonstrated \cite{GaudioTWC2020,KuranageTWC2024} that \ac{AFDM} and \ac{OTFS} can significantly outperform \ac{OFDM} under \ac{DD} conditions.
It {be shown here}, however, that \acp{SIM} can significantly lower the performance gap between these waveforms which in turn, given the potential of the technology to reduce hardware complexity compared to traditional digital signal processing techniques, suggests that the design of waveforms to combat \ac{DD} distortion is can be significantly impacted by the emergence of \acp{SIM}.

From a receive{r} design viewpoint, we aim to estimate the transmit signal $\mathbf{x}$, under the assumption that the effective channel matrix $\bar{\mathbf{H}}$ is known{\footnote{The channel estimation problem -- or equivalently, the sensing problem -- requires further work due to the concatenated nature of the model, specially in cases with \acp{RIS} in the surrounding. 
We therefore relegate this discussion to future work, with some possible directions already discussed in \cite{IimoriTWC2022,ItoOJCOMS2024,TakahashiTWC2024,KuranageTWC2024}.
However, if the channel $\bar{\mathbf{H}}$ is estimated with errors, this error either 1) (if known) can be incorporated when deriving the message passing rules or b) if unknown, approximated and cancelled during the \ac{GaBP} procedure.}}, such that in order to derive a \ac{GaBP}-based detector for arbitrary waveforms, we first consider the generic \ac{I/O} relationship
\vspace{-1ex}
\begin{equation}
\label{General_I/O_arbitrary}
\mathbf{y} = \bar{\mathbf{H}}  \mathbf{x} + \bar{\mathbf{w}}, 
\end{equation}
where, for conciseness, the waveform specific subscripts are dropped from the notation of the effective channel matrix $\bar{\mathbf{H}}$, which for \ac{OFDM}, \ac{OTFS}, and \ac{AFDM} are respectively given by \eqref{eq:OFDM_effective_channel}, \eqref{eq:OTFS_effective_channel} and \eqref{eq:AFDM_effective_channel}.

Setting $\bar{N} = \bar{M} \triangleq N d_s \times N d_s$ with $\bar{n} \triangleq \{1,\dots,\bar{N}\}$ and $\bar{m} \triangleq \{1,\dots,\bar{M}\}$, the element-wise relationship corresponding to equation \eqref{General_I/O_arbitrary} is given by
\begin{equation}
\label{General_I/O_arbitrary_elementwise}
y_{\bar{n}} = \sum_{\bar{m}=1}^{\bar{M}} \bar{h}_{\bar{n},\bar{m}} x_{\bar{m}} + \bar{w}_{\bar{n}}, 
\end{equation}
such that the soft replica of the $\bar{m}$-th communication symbol associated with the $\bar{n}$-th receive signal $y_{\bar{n}}$, computed at the $i$-th iteration of a message-passing algorithm can be denoted by $\hat{x}_{\bar{n},\bar{m}}^{(i)}$, with the corresponding \ac{MSE} of these estimates computed for the $i$-th iteration given by
\begin{equation}
\hat{\sigma}^{2(i)}_{x:{\bar{n},\bar{m}}} \triangleq \mathbb{E}_{x} \big[ | x - \hat{x}_{\bar{n},\bar{m}}^{(i-1)} |^2 \big]= E_\mathrm{S} - |\hat{x}_{\bar{n},\bar{m}}^{(i-1)}|^2, \forall (\bar{n},\bar{m}),
\label{eq:MSE_d_k}
\end{equation}
where $\mathbb{E}_{x}$ refers to expectation over all the possible symbols in the constellation $\mathcal{C}$.

The \ac{GaBP} receiver for such a setup consists of three major stages described below.

\subsubsection{Soft Interference Cancellation} The objective of the \ac{sIC} stage at a given $i$-th iteration of the algorithm is to utilize the soft replicas $\hat{x}_{\bar{n},\bar{m}}^{(i-1)}$ from a previous iteration in order to calculate the data-centric \ac{sIC} signals $\tilde{y}_{x:\bar{n},\bar{m}}^{(i)}$.
Exploiting equation \eqref{General_I/O_arbitrary_elementwise}, such the \ac{sIC} signals are given by
\begin{align}
\label{eq:d_soft_IC}
\tilde{y}_{x:\bar{n},\bar{m}}^{(i)} &= y_{\bar{n}} - \sum_{e \neq \bar{m}} h_{\bar{n},e} \hat{x}_{\bar{n},e}^{(i)}, \\
&= h_{\bar{n},\bar{m}} x_{\bar{m}} + \underbrace{\sum_{e \neq \bar{m}} h_{\bar{n},e}(x_e - \hat{x}_{\bar{n},e}^{(i)}) + \bar{w}_{\bar{n}}}_\text{interference + noise term},
\end{align}

Leveraging the \ac{SGA}, the interference and noise terms in the latter equation can be approximated as Gaussian noise, such that the conditional \acp{PDF} of the \ac{sIC} signals become
\begin{equation}
\label{eq:cond_PDF_d}
\!\!p_{\tilde{\mathrm{y}}_{\mathrm{x}:\bar{n},\bar{m}}^{(i)} \mid \mathrm{x}_{\bar{m}}}(\tilde{y}_{x:\bar{n},\bar{m}}^{(i)}|x_{\bar{m}}) \propto \mathrm{exp}\bigg[ -\frac{|\tilde{y}_{x:\bar{n},\bar{m}}^{(i)}\! -\! h_{\bar{n},\bar{m}} x_{\bar{m}}|^2}{\tilde{\sigma}_{x:\bar{n},\bar{m}}^{2(i)}} \bigg]\!,
\end{equation}
with their conditional variances expressed as
\begin{equation}
\label{eq:soft_IC_var_d}
\tilde{\sigma}_{x:\bar{n},\bar{m}}^{2(i)} = \sum_{e \neq \bar{m}} \left|h_{\bar{n},e}\right|^2 \hat{\sigma}^{2(i)}_{x:{\bar{n},e}} + \sigma^2_w.
\end{equation}


\subsubsection{Belief Generation} In the belief generation stage of the algorithm the \ac{SGA} is exploited under the assumptions that $\bar{N}$ is a sufficiently large number, and that the individual estimation errors in $\hat{x}_{\bar{n},\bar{m}}^{(i-1)}$ are independent, in order to generate initial estimates (aka beliefs) for all the data symbols.

\begin{algorithm}[H]
\caption{SIM Optimization and Data Detection}
\label{alg:proposed_decoder}
\setlength{\baselineskip}{11pt}
\textbf{Input:} receive signal vector $\mathbf{y}\in\mathbb{C}^{\bar{N}\times 1}$, complex channel matrix $\bar{\mathbf{H}} \in \mathbb{C}^{\bar{N}\times \bar{M}}$, number of \ac{GaBP} iterations $i_{\max}$, number of gradient descent iterations $i_{\mathrm{GD}}$, data constellation power $E_\mathrm{S}$, noise variance $\sigma^2_w$ and damping factor $\beta_x$. \\
\textbf{Output:} $\hat{\mathbf{x}}$ 
\vspace{-2ex} 
\begin{algorithmic}[1]  
\STATEx \hspace{-3.5ex}\hrulefill
\STATEx \hspace{-3.5ex}\textbf{Initialization}
\STATEx \hspace{-3.5ex} - Set iteration counter to $i=0$ and amplitudes $c_x = \sqrt{E_\mathrm{S}/2}$.
\STATEx \hspace{-3.5ex} - Set initial data estimates to $\hat{x}_{\bar{n},\bar{m}}^{(0)} = 0$ and corresponding 
\STATEx \hspace{-2ex} variances to $\hat{\sigma}^{2(0)}_{x:{\bar{n},\bar{m}}} = E_\mathrm{S}, \forall \bar{n},\bar{m}$.
\STATEx \hspace{-3.5ex}\hrulefill
\STATEx \hspace{-3.5ex}\textbf{Steepest Ascent-based SIM Optimization}
\STATEx \hspace{-3.5ex}\textbf{for} $i=1$ to $i_{\mathrm{GD}}$ \textbf{do}: $\forall q, \tilde{q}, m, \tilde{m}$
\STATE Compute the gradients from equations \eqref{eq:final_grad_TXSIM} and \eqref{eq:final_grad_RXSIM}.
\STATE Update normalization parameters from equation \eqref{eq:normalization_params}.
\STATE Update the phase parameters via equation \eqref{eq:param_updates_GD}.

\STATEx \hspace{-3.5ex}\textbf{end for}
\STATEx \hspace{-3.5ex}\textbf{GaBP-based Data Detection}
\STATEx \hspace{-3.5ex}\textbf{for} $i=1$ to $i_\text{max}$ \textbf{do}: $\forall \bar{n}, \bar{m}$
\STATE Compute \ac{sIC} data signal $\tilde{y}_{x:{\bar{n},\bar{m}}}^{(i)}$ and its corresponding variance $\tilde{\sigma}^{2(i)}_{x:{\bar{n},\bar{m}}}$ from equations \eqref{eq:d_soft_IC} and \eqref{eq:soft_IC_var_d}.
\STATE Compute extrinsic data signal belief $\bar{x}_{\bar{n},\bar{m}}^{(i)}$ and its corresponding variance $\bar{\sigma}_{x:{\bar{n},\bar{m}}}^{2(i)}$ from equations \eqref{eq:extrinsic_mean_d} and \eqref{eq:extrinsic_var_d}.
\STATE Compute denoised and damped data signal estimate $\hat{x}_{\bar{n},\bar{m}}^{(i)}$ from equations \eqref{eq:QPSK_denoiser} and \eqref{eq:d_damped}.
\STATE Compute denoised and damped data signal variance $\hat{\sigma}_{x:{\bar{n},\bar{m}}}^{2(i)}$ from equations \eqref{eq:MSE_d_k} and \eqref{eq:MSE_d_m_damped}.

\STATEx \hspace{-3.5ex}\textbf{end for}
\STATE Calculate $\hat{x}_{\bar{m}}, \forall \bar{m}$ (equivalently $\hat{\mathbf{x}}$) using equation \eqref{eq:d_hat_final_est}. 

\end{algorithmic}
\end{algorithm}

As a consequence of the \ac{SGA} and with the conditional \acp{PDF} of equation \eqref{eq:cond_PDF_d}, the following extrinsic \acp{PDF}
\vspace{-1ex}
\begin{equation}
\label{eq:extrinsic_PDF_d}
\prod_{e \neq \bar{n}} p_{\tilde{\mathrm{y}}_{\mathrm{x}:e,\bar{m}}^{(i)} \mid \mathrm{x}_{\bar{m}}}(\tilde{y}_{x:e,\bar{m}}^{(i)}|x_{\bar{m}}) \propto \mathrm{exp}\bigg[ - \frac{(x_{\bar{m}} - \bar{x}_{\bar{n},\bar{m}}^{(i)})^2}{\bar{\sigma}_{x:\bar{n},\bar{m}}^{2(i)}} \bigg],
\end{equation}
are obtained, where the corresponding extrinsic means and variances are respectively defined as
\vspace{-1ex}
\begin{equation}
\label{eq:extrinsic_mean_d}
\bar{x}_{\bar{n},\bar{m}}^{(i)} = \bar{\sigma}_{x:\bar{n},\bar{m}}^{(i)} \sum_{e \neq \bar{n}} \frac{h^*_{e,\bar{m}} \tilde{y}_{x:e,\bar{m}}^{(i)}}{ \tilde{\sigma}_{x:e,\bar{m}}^{2(i)}},
\end{equation}
\vspace{-1ex}
\begin{equation}
\label{eq:extrinsic_var_d}
\bar{\sigma}_{x:\bar{n},\bar{m}}^{2(i)} = \bigg( \sum_{e \neq \bar{n}} \frac{|h_{e,\bar{m}}|^2}{\tilde{\sigma}_{x:e,\bar{m}}^{2(i)}} \bigg)^{\!\!\!-1},
\end{equation}
with $h^*_{e,\bar{m}}$ denoting the complex conjugate of $h_{e,\bar{m}}$.

\subsubsection{Soft Replica Generation} Finally, the soft replica generation stage consists of denoising the previously computed beliefs under a Bayes-optimal rule, in order to obtain the final estimates for the desired variables.
For \ac{QPSK} modulation\footnote{We consider \ac{QPSK} for simplicity, but \ac{wlg}, since denoisers for other modulation schemes can also be designed \cite{TakahashiTCOM2019}.}, the Bayes-optimal denoiser is given by
\vspace{-1ex}
\begin{equation}
\hat{x}_{\bar{n},\bar{m}}^{(i)}\! =\! c_x \bigg(\! \text{tanh}\!\bigg[ 2c_d \frac{\Real{\bar{x}_{\bar{n},\bar{m}}^{(i)}}}{\bar{\sigma}_{x:{\bar{n},\bar{m}}}^{2(i)}} \bigg]\!\! +\! \jmath \text{tanh}\!\bigg[ 2c_d \frac{\Imag{\bar{x}_{\bar{n},\bar{k}}^{(i)}}}{\bar{\sigma}_{{x}:{\bar{n},\bar{m}}}^{2(i)}} \bigg]\!\bigg),\!\!
\label{eq:QPSK_denoiser}
\end{equation}
where $c_x \triangleq \sqrt{E_\mathrm{S}/2}$ denotes the magnitude of the real and imaginary parts of the explicitly chosen \ac{QPSK} symbols, with its corresponding variance updated as in equation \eqref{eq:MSE_d_k}.
\newpage

After obtaining $\hat{x}_{\bar{n},\bar{m}}^{(i)}$ as per equation \eqref{eq:QPSK_denoiser}, the final outputs are computed by damping the results to prevent convergence to local minima due to incorrect hard-decision replicas \cite{Su_TSP_2015}.
Letting the damping factor be $0 < \beta_x < 1$ yields
\vspace{-1ex}
\begin{equation}
  \vspace{-1ex}
\label{eq:d_damped}
\hat{x}_{\bar{n},\bar{m}}^{(i)} = \beta_x \hat{x}_{\bar{n},\bar{m}}^{(i)} + (1 - \beta_x) \hat{x}_{\bar{n},\bar{m}}^{(i-1)}.
\end{equation}

Similarly, the variances $\hat{\sigma}^{2(i)}_{x:{\bar{n},\bar{m}}}$ are first updated via equation \eqref{eq:MSE_d_k} and then damped via
\vspace{-1ex}
\begin{equation}
  \vspace{-1ex}
\label{eq:MSE_d_m_damped}
\hat{\sigma}^{2(i)}_{x:{\bar{n},\bar{m}}} = \beta_x \hat{\sigma}_{x:{\bar{n},\bar{m}}}^{2(i)} + (1-\beta_x) \hat{\sigma}_{x:{\bar{n},\bar{m}}}^{2(i-1)},
\end{equation}

Finally, as a result of the conflicting dimensions, the consensus update of the estimates can be obtained as
\vspace{-1ex}
\begin{equation}
  \vspace{-1ex}
\label{eq:d_hat_final_est}
\hat{x}_{\bar{m}} = \bigg( \sum_{\bar{n}=1}^{\bar{N}} \frac{|h_{\bar{n},\bar{m}}|^2}{\tilde{\sigma}_{x:\bar{n},\bar{m}}^{2(i_\text{max})}} \bigg)^{\!\!\!-1} \! \! \bigg( \sum_{\bar{n}=1}^{\bar{N}} \frac{h^*_{\bar{n},\bar{m}} \tilde{y}_{x:\bar{n},\bar{m}}^{(i_\text{max})}}{ \tilde{\sigma}_{x:\bar{n},\bar{m}}^{2(i_\text{max})}} \bigg).
\end{equation}

The complete pseudocode for the \ac{SIM} parametrization and detection procedure here proposed is summarized above in Algorithm \ref{alg:proposed_decoder}.

\vspace{-1ex}
{
\subsection{Complexity Analysis}
\vspace{-1ex}

A major component of the computational cost of solving the optimization problem \eqref{eq:full_optimization_problem_channel_coeff} is the iterative computation of the sub-gradients given by equations \eqref{eq:final_grad_TXSIM} and \eqref{eq:final_grad_RXSIM}, as described in line 1 of Algorithm \ref{alg:proposed_decoder}.
To that end, in order to obtain a tractable relationship and under the approximations $M \approx \tilde{M}$, $Q \approx \tilde{Q}$ and $N_\mathrm{T} \approx N_\mathrm{R}$ (with $M >> Q$ and $M >> N_\mathrm{T}$), a simplified complete computational complexity expression can be extracted as $\mathcal{O}\big( PM^2 N_\mathrm{T} + Q^2M^3 + PQN_\mathrm{T}M^2 \big)$, where the first term is due to executing equations \eqref{eq:transmit_SIM_full} and \eqref{eq:receive_SIM_full}, the second term is due to the multiplications in the gradient expressions defined in equation \eqref{eq:Upsilon_t} and \eqref{eq:Upsilon_r} and the last term is due to executing equations \eqref{S_TX_part} and \eqref{S_RX_part} in Algorithm \ref{alg:proposed_decoder}.
As can be seen from the expression, the computational complexity of the proposed gradient ascent method is dependent on the total number of paths, the size of the \ac{TX} and \ac{RX} antenna arrays, the number of \ac{TX} and \ac{RX} \ac{SIM} layers and the number of meta-atoms on each layer.

On the other hand, the complexity of the proposed \ac{GaBP} detection algorithm is linear on the number of element-wise operations, and its per-iteration computational complexity is given by $\mathcal{O}(\bar{N}\bar{M})$. 
Notice that this complexity is much lesser than that of typical detection methods such as the \ac{LMMSE}, which is $\mathcal{O}(\bar{N}^3)$ due to the costly matrix inversion involved.

}

\vspace{-2ex}
\subsection{Performance Analysis}

Finally, we utilize the aforementioned methods to compare via simulations the \ac{BER} performances of \ac{OFDM}, \ac{OTFS} and \ac{AFDM} systems with \ac{QPSK} modulation in a \ac{SIM}-enabled \ac{DD} channel.
For the sake of simplicity, we consider uplink \ac{SIMO} and \ac{SISO} scenarios\footnote{Notice that the generalization to a full \ac{MIMO} case would require the design of optimum \ac{TX}-\ac{BF} and a {receiver that can} deal with the resulting spatial correlation, {both of which are non-trivial and therefore} will be pursued in a follow-up work.} where both the \ac{TX} and \ac{RX} are equipped with \acp{SIM}, and there are no \acp{RIS} in the environment.
All the other parameters are as in section \ref{subsec:comp_eff_channel} with the exception that the number of paths here is set to a more realistic $P=5$ and that the delay and Doppler shifts are randomly generated.

\vspace{-1ex}
\begin{figure}[H]
\centering
\captionsetup[subfloat]{labelfont=small,textfont=small}
\subfloat[SIMO: $N_\mathrm{R} = 4$.]{{\includegraphics[width=\columnwidth]{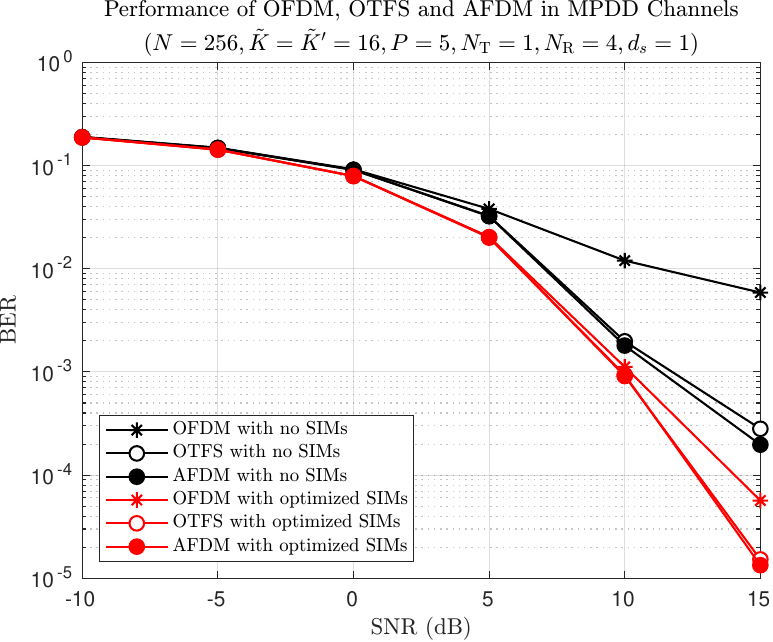}}}%
\label{fig:BER_SIM_SIMO_4}
\par\bigskip
\vspace{-1ex}
\subfloat[SISO: $N_\mathrm{R} = 1$.]{{\includegraphics[width=\columnwidth]{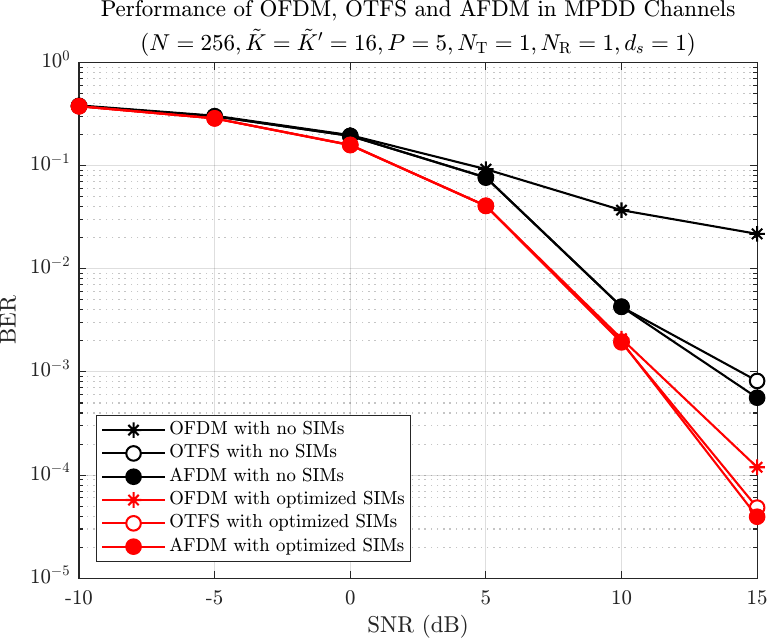}}}%
\label{fig:BER_SIM_SISO}
\caption{\ac{BER} Performance of \ac{OFDM}, \ac{OTFS} and \ac{AFDM} waveforms with \ac{QPSK} modulation in \ac{MPDD} channels with high-mobility, with \acp{SIM} placed at very close distances to both the \ac{TX} and the \ac{RX} {for $Q = \tilde{Q} = 5$ and $M = \tilde{M} = 100$.}}
\label{fig:BER_SIM}
\vspace{-2ex}
\end{figure}

{
In order to make sure that no power advantage other than the passive \ac{SIM} gains resulting from the \ac{SIM} parametrization results, we enforce that the complete effective channels have identical power such that $||\bar{\mathbf{H}}_\text{OFDM}||^2_F = ||\bar{\mathbf{H}}_\text{OTFS}||^2_F = ||\bar{\mathbf{H}}_\text{AFDM}||^2_F = ||\bar{\mathbf{H}}_\text{MIMO}||^2_F$ for all the cases in the comparison.
Notice that this is done to the \ul{disadvantage} of our contribution, which is therefore only highlighted given the performance improvements observed.
}

The results are shown in Figure \ref{fig:BER_SIM}, from which it can be seen that the \ac{SIM}-based systems significantly outperforms those without \acp{SIM}.

It can be observed that other than the obvious reduction of \ac{BER} resulting from the utilization of more receive antennas, the systems employing \acp{SIM} exhibit very similar behaviors, which indicates that the technology indeed has the potential to substantially lower the gap between more sophisticated (multi-antenna) and less sophisticated (single-antenna) systems.
In fact, it can also be observed that having \acp{SIM} significantly boosts the performance of conventional \ac{OFDM} compared to its typical performance in a \ac{DD} channel, bringing it close to that of far more sophisticated schemes such as \ac{OTFS} and \ac{AFDM}.
{
However, also notice that while the \ac{SIMO} case shows a clear performance gain, there is a slight deviation from classical \ac{MIMO} theory due to the message-passing inefficiencies caused as a consequence of the correlated nature of the effective \ac{DD} channel $\bar{\mathbf{H}}$ in \ac{SIMO}/\ac{MIMO} cases\footnotemark.

}

\begin{figure}[H]
\centering
{{\includegraphics[width=\columnwidth]{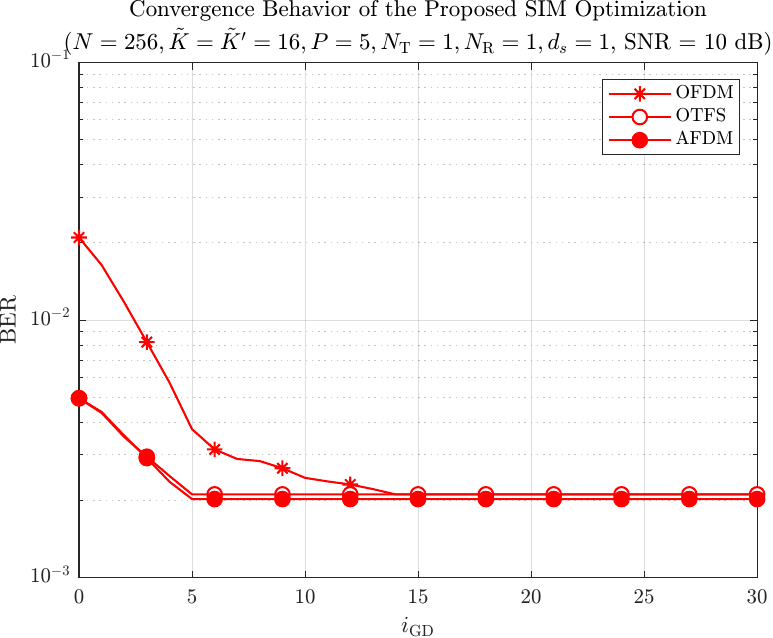}}}
\caption{Convergence behavior of \ac{OFDM}, \ac{OTFS} and \ac{AFDM} waveforms with \ac{QPSK} modulation in \ac{MPDD} channels with high-mobility, with \ac{SIM} placed at very close distances to both the \ac{TX} and the \ac{RX} in a \ac{SISO} setting.}
\label{fig:BER_SIM_SISO_conv}
\vspace{1ex}
{{\includegraphics[width=\columnwidth]{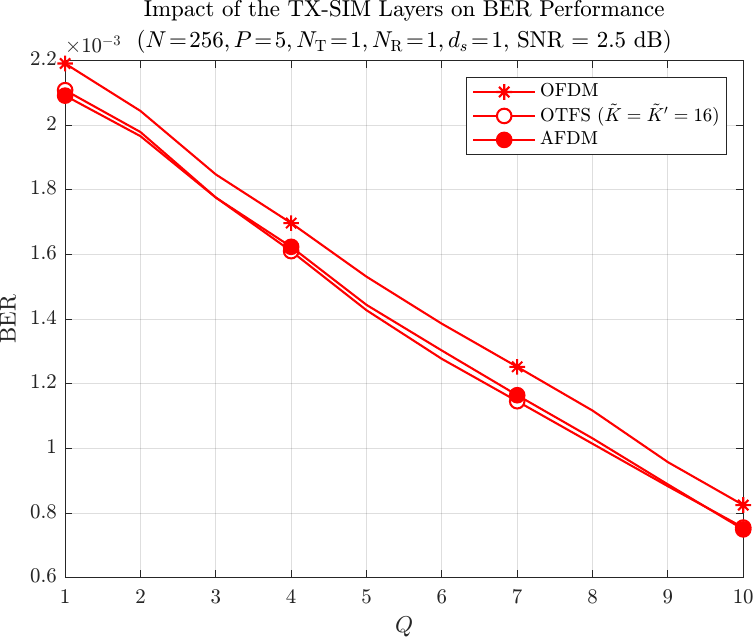}}}
\caption{BER Performance vs. changing TX-SIM layers of \ac{OFDM}, \ac{OTFS} and \ac{AFDM} waveforms with \ac{QPSK} modulation in \ac{MPDD} channels with high-mobility, with \ac{SIM} placed at very close distances to both the \ac{TX} and the \ac{RX} in a \ac{SISO} setting.}
\label{fig:BER_SIM_SISO_TX_layers}
\vspace{-2ex}
\end{figure}

\begin{figure}[H]
\centering
{{\includegraphics[width=\columnwidth]{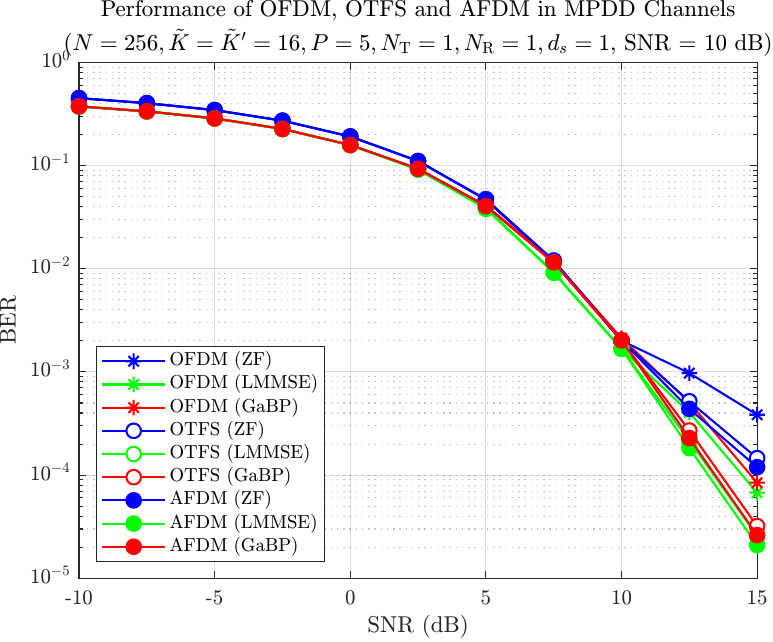}}}
\caption{BER Performance for various detectors using \ac{OFDM}, \ac{OTFS} and \ac{AFDM} waveforms with \ac{QPSK} modulation in \ac{MPDD} channels with high-mobility, with \ac{SIM} placed at very close distances to both the \ac{TX} and the \ac{RX} in a \ac{SISO} setting.}
\label{fig:BER_SIM_SISO_detectors}
\end{figure}

\footnotetext{A potential solution to this would be to incorporate a \ac{PDA}-based receiver as done in \cite{KuranageTWC2024,Ranasinghe_BISTATIC_SIM_ARXIV_2025} to mitigate the effects of the correlations.}

{
In order to illustrate that the proposed steepest-ascent-based \ac{SIM} optimization converges, we also show convergence results in Fig. \ref{fig:BER_SIM_SISO_conv} for the \ac{SISO} case.
As seen from the results, the \ac{SIM} optimization converges to a steady state after a few iterations, and the \ac{BER} performance of the system improves as the optimization progresses.

Next, we switch focus to analyze the \ac{BER} performance of the \ac{SISO} case where the number of \ac{TX}-\ac{SIM} layers $Q$ vary.
For a fair comparison leveraging the full abilities of the \acp{SIM}, we relax the normalization conditions imposed in the latter part of this subsection for this particular result.
As seen from the results in Fig. \ref{fig:BER_SIM_SISO_TX_layers}, while we only transmit a single stream of data for this \ac{SISO} case, \ac{BER} performance improves with an increasing number of \ac{SIM} layers which is in agreement with the ``lensing '' effect achieved via \acp{SIM}.

Finally, we analyze the \ac{BER} performance of the \ac{SISO} case with different detectors in Fig. \ref{fig:BER_SIM_SISO_detectors}.
In conjunction with the low-complexity \ac{GaBP} detector presented in Section \ref{subsec:GaBP}, we also show the performance of the \ac{LMMSE} and \ac{ZF} detectors, which are the most common detectors used in communications systems.
For clarity, for the model showcased in equation \eqref{General_I/O_arbitrary}, we denote the \ac{LMMSE} and \ac{ZF} detectors as $\hat{\mathbf{x}}_\text{LMMSE}$ and $\hat{\mathbf{x}}_\text{ZF}$ respectively, and define them as
\begin{subequations}
\label{eq:detectors}
\begin{equation}
\label{eq:LMMSE_detector}
\hat{\mathbf{x}}_\text{LMMSE} = \left( \bar{\mathbf{H}}^H \bar{\mathbf{H}} + \sigma^2_w \mathbf{I}_{\bar{N}} \right)^{-1} \bar{\mathbf{H}}^H \mathbf{y},
\end{equation}
\begin{equation}
\label{eq:ZF_detector}
\hat{\mathbf{x}}_\text{ZF} = \left( \bar{\mathbf{H}}^H \bar{\mathbf{H}} \right)^{-1} \bar{\mathbf{H}}^H \mathbf{y}.
\end{equation}
\end{subequations}

As seen from the results in Fig. \ref{fig:BER_SIM_SISO_detectors}, the \ac{GaBP} detector outperforms the \ac{ZF} detector at the high \ac{SNR} regime, and is very close to the performance of the \ac{LMMSE} detector, at a much lower complexity than either of them due to the lack of costly matrix inversion operations.

}

\section{Conclusion}

We introduced a novel \acl{MPDD} \ac{MIMO} channel model that incorporates an arbitrary number of \acp{RIS} in the ambient, as well as \acp{SIM} equipping both the transmitter and receiver, which can be applied to various \ac{ISAC}-enabling waveforms.
Offering explicit \ac{I/O} relationships for \ac{OFDM}, \ac{OTFS}, and \ac{AFDM} in particular, we then augmented this discussion by showing how the proposed \ac{MPDD} channel model can be seamlessly applied to optimize \acp{SIM} in order to improve the performance of these waveforms in \ac{DD} environments. 
The results indicate that \acp{SIM} positively and significantly impact the performances of such systems, having the potential to reduce the gap between recent and more sophisticated waveform design approaches, such as \ac{OTFS} and \ac{AFDM}, and the classic \ac{OFDM}.
A further study on the impact of \acp{SIM} onto the sensing performance of \ac{ISAC} systems, and a full data detection algorithm capable of handling the correlations in the effective \ac{MIMO} channel will be carried out in a follow-up work.
{In addition, further work is required to study the impact of \acp{RIS} in the environment, and how to jointly optimize the \ac{TX}-\ac{BF} and \ac{RX}-\ac{BF} with the \acp{SIM} in order to maximize the performance of the system.}

\bibliographystyle{IEEEtran}

\begin{thebibliography}{10}
\providecommand{\url}[1]{#1}
\csname url@samestyle\endcsname
\providecommand{\newblock}{\relax}
\providecommand{\bibinfo}[2]{#2}
\providecommand{\BIBentrySTDinterwordspacing}{\spaceskip=0pt\relax}
\providecommand{\BIBentryALTinterwordstretchfactor}{4}
\providecommand{\BIBentryALTinterwordspacing}{\spaceskip=\fontdimen2\font plus
\BIBentryALTinterwordstretchfactor\fontdimen3\font minus
  \fontdimen4\font\relax}
\providecommand{\BIBforeignlanguage}[2]{{%
\expandafter\ifx\csname l@#1\endcsname\relax
\typeout{** WARNING: IEEEtran.bst: No hyphenation pattern has been}%
\typeout{** loaded for the language `#1'. Using the pattern for}%
\typeout{** the default language instead.}%
\else
\language=\csname l@#1\endcsname
\fi
#2}}
\providecommand{\BIBdecl}{\relax}
\BIBdecl

\bibitem{ChenCSM2017}
S.~Chen, J.~Hu, Y.~Shi, Y.~Peng, J.~Fang, R.~Zhao, and L.~Zhao,
  ``Vehicle-to-Everything ({V2X}) Services Supported by {LTE}-Based Systems and
  {5G},'' \emph{IEEE Commun. Stand. Mag.}, vol.~1, no.~2, 2017.

\bibitem{NguyenIoTJ2022}
D.~C. Nguyen, M.~Ding, P.~N. Pathirana, A.~Seneviratne, J.~Li, D.~Niyato,
  O.~Dobre, and H.~V. Poor, ``{6G} Internet of Things: A Comprehensive
  Survey,'' \emph{IEEE Internet Things J.}, vol.~9, no.~1, 2022.

\bibitem{ShiNetwork2024}
J.~Shi, Z.~Li, J.~Hu, Z.~Tie, S.~Li, W.~Liang, and Z.~Ding, ``{OTFS} Enabled
  {LEO} Satellite Communications: {A} Promising Solution to Severe {D}oppler
  Effects,'' \emph{IEEE Network}, vol.~38, no.~1, 2024.

\bibitem{LiCOMMST2022}
J.~Li, Y.~Niu, H.~Wu, B.~Ai, S.~Chen, Z.~Feng, Z.~Zhong, and N.~Wang,
  ``Mobility Support for Millimeter Wave Communications: Opportunities and
  Challenges,'' \emph{IEEE Commun. Surveys Tuts.}, vol.~24, no.~3, 2022.

\bibitem{WangTWC2006}
T.~Wang, J.~Proakis, E.~Masry, and J.~Zeidler, ``Performance Degradation of
  {OFDM} Systems due to Doppler Spreading,'' \emph{IEEE Trans. Wireless
  Commun.}, vol.~5, no.~6, 2006.

\bibitem{LiuTIT2004}
K.~Liu, T.~Kadous, and A.~Sayeed, ``Orthogonal Time-Frequency Signaling over
  Doubly Dispersive Channels,'' \emph{IEEE Trans. Inf. Theory}, vol.~50,
  no.~11, 2004.

\bibitem{Bliss_Govindasamy_2013}
  D.~W. Bliss and S.~Govindasamy, \emph{Dispersive and Doubly Dispersive
    Channels}, Cambridge University Press,
    2013.

\bibitem{Rou_SPM_2024}
H.~S. Rou, G.~T.~F. de~Abreu, J.~Choi, D.~González~G., M.~Kountouris, Y.~L.
  Guan, and O.~Gonsa, ``From Orthogonal Time–Frequency Space to Affine
  Frequency-Division Multiplexing: A Comparative Study of Next-Generation
  Waveforms for Integrated Sensing and Communications in Doubly Dispersive
  Channels,'' \emph{IEEE Signal Process. Mag.}, vol.~41, no.~5, 2024.

\bibitem{SurabhiTWC2019}
G.~D. Surabhi, R.~M. Augustine, and A.~Chockalingam, ``On the Diversity of
  Uncoded {OTFS} Modulation in Doubly-Dispersive Channels,'' \emph{IEEE Trans.
  Wireless Commun.}, vol.~18, no.~6, 2019.

\bibitem{BomfinTWC2021}
R.~Bomfin, M.~Chafii, A.~Nimr, and G.~Fettweis, ``A Robust Baseband Transceiver
  Design for Doubly-Dispersive Channels,'' \emph{IEEE Trans. Wireless Commun.},
  vol.~20, no.~8, 2021.

\bibitem{PfadlerTWC2024}
A.~Pfadler, T.~Szollmann, P.~Jung, and S.~Stańczak, ``Estimation of
  Doubly-Dispersive Channels in Linearly Precoded Multicarrier Systems using
  Smoothness Regularization,'' \emph{IEEE Trans. Wireless Commun.}, vol.~23,
  no.~2, 2024.

\bibitem{LiangTWC2024}
Y.~Liang, P.~Fan, Q.~Wang, and X.~He, ``Two-Dimensional Delay-Doppler Pilots
  and Channel Estimation for Multi-Antenna {OTFS} in Doubly Dispersive
  Channels,'' \emph{IEEE Trans. Wireless Commun.}, vol.~23, no.~7, 2024.

\bibitem{HaifTVT2024}
H.~Haif, S.~E. Zegrar, and H.~Arslan, ``Novel {OCDM} Transceiver Design for
  Doubly-Dispersive Channels,'' \emph{IEEE Trans. Veh. Technol.}, vol.~73,
  no.~8, 2024.

\bibitem{ZhangJSAC2024}
H.~Zhang, S.~Chen, W.~Meng, J.~Yuan, and C.~Li, ``Multiuser Association and
  Localization over Doubly Dispersive Multipath Channels for Integrated Sensing
  and Communications,'' \emph{IEEE J. Sel. Areas Commun.}, vol.~42, no.~10, 2024.

\bibitem{Liu_JSAC_2022}
F.~Liu, Y.~Cui, C.~Masouros, J.~Xu, T.~X. Han, Y.~C. Eldar, and S.~Buzzi,
  ``Integrated {S}ensing and {C}ommunications: {T}oward {D}ual-{F}unctional
  {W}ireless {N}etworks for {6G} and {B}eyond,'' \emph{IEEE J. Sel. Areas
  Commun.}, vol.~40, no.~6, 2022.

\bibitem{CSL2023}
S.~P. Chepuri, N.~Shlezinger, F.~Liu, G.~C. Alexandropoulos, S.~Buzzi, and
  Y.~C. Eldar, ``Integrated Sensing and Communications with Reconfigurable
  Intelligent Surfaces: From Signal Modeling to Processing,'' \emph{IEEE Signal
  Process. Mag.}, vol.~40, no.~6, Sep. 2023.

\bibitem{GonzalezProcIEEE2024}
N.~González-Prelcic, M.~F. Keskin, O.~Kaltiokallio, M.~Valkama, D.~Dardari,
  X.~Shen, Y.~Shen, M.~Bayraktar, and H.~Wymeersch, ``The Integrated Sensing
  and Communication Revolution for {6G}: Vision, Techniques, and
  Applications,'' \emph{Proc. IEEE}, early access, 2024.

\bibitem{FD_MIMO_ISAC_2024}
B.~Smida, G.~C. Alexandropoulos, T.~Riihonen, and M.~A. Islam, ``In-Band
  Full-Duplex {MIMO} Systems for Simultaneous Communications and Sensing:
  Challenges, Methods, and Future Perspectives,'' \emph{IEEE Signal Process.
  Mag.}, vol.~41, no.~5, Sep. 2024.

\bibitem{HyeonTWC2024}
H.~S. Rou, G.~T.~F. de~Abreu, D.~González~G., and O.~Gonsa, ``Integrated
  Sensing and Communications for {3D} Object Imaging via Bilinear Inference,''
  \emph{IEEE Trans. Wireless Commun.}, vol.~23, no.~8, 2024.

\bibitem{GaudioTWC2020}
L.~Gaudio, M.~Kobayashi, G.~Caire, and G.~Colavolpe, ``On the Effectiveness of
  {OTFS} for Joint Radar Parameter Estimation and Communication,'' \emph{IEEE
  Trans. Wireless Commun.}, vol.~19, no.~9, 2020.

\bibitem{Mohammed_BITS_2022}
S.~K. Mohammed, R.~Hadani, A.~Chockalingam, and R.~Calderbank, ``{OTFS}---{A}
  Mathematical Foundation for Communication and Radar Sensing in the
  Delay-Doppler Domain,'' \emph{IEEE Inf. Theory Mag.}, vol.~2, no.~2, 2022.

\bibitem{Gupta_OJCS_2024}
A.~Gupta, M.~Jafri, S.~Srivastava, A.~K. Jagannatham, and L.~Hanzo, ``An Affine
  Precoded Superimposed Pilot based mm{W}ave {MIMO}-{OFDM} {ISAC} System,''
  \emph{IEEE Open J. Commun. Soc.}, vol.~5, 2024.

\bibitem{Ranasinghe_ICASSP_2024}
K.~R.~R. Ranasinghe, H.~S. Rou, and G.~T.~F. de~Abreu, ``Fast and Efficient
  Sequential Radar Parameter Estimation in {MIMO}-{OTFS} Systems,'' in
  \emph{Proc. IEEE International Conference on Acoustics, Speech and Signal Processing (ICASSP)}, Seoul, South Korea, 2024.

\bibitem{KuranageTWC2024}
K.~R.~R. Ranasinghe, H.~S. Rou, G.~T.~F. De~Abreu, T.~Takahashi, and K.~Ito,
  ``Joint Channel, Data and Radar Parameter Estimation for {AFDM} Systems in
  Doubly-Dispersive Channels,'' \emph{IEEE Trans. Wireless Commun.}, early access, 2024.

\bibitem{Bemani_WCL_2024}
A.~Bemani, N.~Ksairi, and M.~Kountouris, ``Integrated Sensing and
  Communications with Affine Frequency Division Multiplexing,'' \emph{IEEE
  Wireless Commun. Lett.}, vol.~13, no.~5, May 2024, 2024.

\bibitem{Parker_TSP_2014}
J.~T. Parker, P.~Schniter, and V.~Cevher, ``Bilinear {G}eneralized
  {A}pproximate {M}essage {P}assing—{P}art {I}: {D}erivation,'' \emph{IEEE
  Trans. Signal Process.}, vol.~62, no.~22, 2014.

\bibitem{IimoriTWC2021}
H.~Iimori, T.~Takahashi, K.~Ishibashi, G.~T.~F. de~Abreu, and W.~Yu,
  ``Grant-Free Access via Bilinear Inference for Cell-Free MIMO with
  Low-Coherence Pilots,'' \emph{IEEE Trans. Wireless Commun.},
  vol.~20, no.~11, 2021.

\bibitem{TakahashiTWC2023}
T.~Takahashi, H.~Iimori, K.~Ando, K.~Ishibashi, S.~Ibi, and G.~T.~F. de~Abreu,
  ``Bayesian Receiver Design via Bilinear Inference for Cell-Free Massive MIMO
  with Low-Resolution ADCs,'' \emph{IEEE Trans. Wireless Commun.}, vol.~22, no.~7, 2023.

\bibitem{YangCommL2011}
X.~Yang, K.~Lei, S.~Peng, and X.~Cao, ``Blind Detection for Primary User based
  on the Sample Covariance Matrix in Cognitive Radio,'' \emph{IEEE
  Commun. Lett.}, vol.~15, no.~1, 2011.

\bibitem{BaoACCESS2019}
J.~Bao, J.~Nie, C.~Liu, B.~Jiang, F.~Zhu, and J.~He, ``Improved Blind Spectrum
  Sensing by Covariance Matrix Cholesky Decomposition and RBF-SVM Decision
  Classification at Low SNRs,'' \emph{IEEE Access}, vol.~7, 2019.

\bibitem{RanasingheWCNC2024}
K.~R.~R. Ranasinghe, K.~Ando, H.~S. Rou, G.~T.~F. de~Abreu, and A.~Bathelt,
  ``Blind Bistatic Radar Parameter Estimation in Doubly-Dispersive Channels,'' to appear in \emph{Proc. IEEE Wireless Communications and Networking Conference (WCNC)}, Milan, Italy, 2024. 

\bibitem{Ni_ISWCS_2022}
Y.~Ni, Z.~Wang, P.~Yuan, and Q.~Huang, ``An {AFDM}-based Integrated Sensing and
  Communications,'' in \emph{Proc. International Symposium on Wireless Communication Systems}, Hangzhou, China, 2022.

\bibitem{Bemani_TWC_2023}
A.~Bemani, N.~Ksairi, and M.~Kountouris, ``Affine Frequency Division
  Multiplexing for Next Generation Wireless Communications,'' \emph{IEEE Trans.
  Wireless Commun.}, vol.~22, no.~11, 2023.

\bibitem{RouAsilimoar2024}
H.~S. Rou, K.~Yukiyoshi, T.~Mikuriya, G.~T.~F. de~Abreu, and N.~Ishikawa,
  ``AFDM Chirp-Permutation-Index Modulation with Quantum-Accelerated Codebook
  Design,'' to appear in \emph{Proc. IEEE 57th Asilomar Conference on Signals, Systems, and Computers (Asilomar CSSC)}, Pacific Grove, CA, USA, 2024.

\bibitem{LiyanaarachchiTWC2024}
S.~D. Liyanaarachchi, T.~Riihonen, C.~B. Barneto, and M.~Valkama, ``Joint MIMO
  Communications and Sensing with Hybrid Beamforming Architecture and OFDM
  Waveform Optimization,'' \emph{IEEE Trans. Wireless Commun.},
  vol.~23, no.~2, 2024.

\bibitem{Gaudio_TWC_2022}
L.~Gaudio, G.~Colavolpe, and G.~Caire, ``{OTFS} vs. {OFDM} in the Presence of
  Sparsity: {A} Fair Comparison,'' \emph{IEEE Trans. Wireless Commun.},
  vol.~21, no.~6, 2022.

\bibitem{Srivastava_ISES2018}
S.~Srivastava and P.~Hobden, ``5GHz Chirp Signal Generator for Broadband FMCW
  Radar Applications,'' in \emph{Proc. IEEE International Symposium on Smart Electronic Systems (iSES)}, 2018.

\bibitem{OuyangTCom2016}
X.~Ouyang and J.~Zhao, ``Orthogonal Chirp Division Multiplexing,'' \emph{IEEE
  Trans. Commun.}, vol.~64, no.~9, 2016.

\bibitem{TongTCom2024}
J.~Tong, J.~Yuan, H.~Lin, and J.~Xi, ``Orthogonal Delay-Doppler Division
  Multiplexing (ODDM) over General Physical Channels,'' \emph{IEEE Trans.
  Commun.}, vol.~72, no.~12, 2024.

\bibitem{XLMIMO_tutorial}
Z.~Wang \emph{et~al.}, ``A Tutorial on Extremely Large-Scale {MIMO} for {6G}:
  {F}undamentals, Signal Processing, and Applications,'' \emph{IEEE Commun.
  Surveys Tuts.}, vol.~26, no.~3, 2024.

\bibitem{NF_beam_tracking}
P.~Gavriilidis and G.~C. Alexandropoulos, ``Near-Field Beam Tracking with
  Extremely Massive Dynamic Metasurface Antennas,'' \emph{arXiv preprint
  arXiv:2406.01488}, 2024.

\bibitem{Tsinghua_RIS_Tutorial}
M.~Jian, G.~C. Alexandropoulos, E.~Basar, C.~Huang, R.~Liu, Y.~Liu, and
  C.~Yuen, ``Reconfigurable Intelligent Surfaces for Wireless Communications:
  {O}verview of Hardware Designs, Channel Models, and Estimation Techniques,''
  \emph{Intell. Converged Netw.}, vol.~3, no.~1, 2022.

\bibitem{BAL2024}
E.~Basar, G.~C. Alexandropoulos, Y.~Liu, Q.~Wu, S.~Jin, C.~Yuen, O.~A. Dobre,
  and R.~Schober, ``Reconfigurable Intelligent Surfaces for {6G}: Emerging
  Hardware Architectures, Applications, and Open Challenges,'' \emph{IEEE Veh.
  Technol. Mag.}, vol.~19, no.~3, 2024.

\bibitem{LiuCommST2021}
Y.~Liu, X.~Liu, X.~Mu, T.~Hou, J.~Xu, M.~Di~Renzo, and N.~Al-Dhahir,
  ``Reconfigurable Intelligent Surfaces: Principles and Opportunities,''
  \emph{IEEE Commun. Surveys Tuts.}, vol.~23, no.~3, 2021.

\bibitem{PG13}
C.~Pfeiffer and A.~Grbic, ``Cascaded Metasurfaces for Complete Phase and
  Polarization Control,'' \emph{Appl. Phys. Lett.}, vol. 102, no.~23, 2013.

\bibitem{ZKW+18}
Y.~Zhou, I.~I. Kravchenko, H.~Wang, J.~R. Nolen, G.~Gu, and J.~Valentine,
  ``Multilayer Noninteracting Dielectric Metasurfaces for Multiwavelength
  Metaoptics,'' \emph{Nano. Lett.}, vol.~18, no.~12, 2018.

\bibitem{HLC+19}
Y.~Hu, X.~Luo, Y.~Chen, Q.~Liu, X.~Li, Y.~Wang, N.~Liu, and H.~Duan,
  ``{3D}-Integrated Metasurfaces for Full-Colour Holography,'' \emph{Light Sci.
  Appl.}, vol.~8, no.~1, 2019.

{
\bibitem{YaoWCL2024}
X.~Yao, J.~An, L.~Gan, M.~Di~Renzo and C.~Yuen, ``Channel Estimation for Stacked Intelligent Metasurface-Assisted Wireless Networks,'' \emph{IEEE
  Wireless Commun. Lett.}, vol.~13, no.~5, May 2024.

\bibitem{NiuWCL2024}
H.~Niu, J.~An, A.~Papazafeiropoulos, L.~Gan, S.~Chatzinotas and M.~Debbah, ``Stacked Intelligent Metasurfaces for Integrated Sensing and Communications,'' \emph{IEEE
  Wireless Commun. Lett.}, vol.~13, no.~10, Oct. 2024.

\bibitem{LiTVT2025}
S.~Li, F.~Zhang, T.~Mao, R.~Na, Z.~Wang and G.~K.~Karagiannidis, ``Transmit Beamforming Design for ISAC With Stacked Intelligent Metasurfaces,'' \emph{IEEE Trans. Veh. Technol.}, vol.~74,
    no.~4, April 2025.
}

\bibitem{SrivaTWC2022}
S.~Srivastava, R.~K. Singh, A.~K. Jagannatham, A.~Chockalingam, and L.~Hanzo,
  ``{OTFS} Transceiver Design and Sparse Doubly-Selective {CSI} Estimation in
  Analog and Hybrid Beamforming aided mmWave {MIMO} Systems,'' \emph{IEEE
  Trans. Wireless Commun.}, vol.~21, no.~12, 2022.

\bibitem{AnWC2024}
J.~An, C.~Yuen, C.~Xu, H.~Li, D.~W.~K. Ng, M.~Di~Renzo, M.~Debbah, and
  L.~Hanzo, ``Stacked Intelligent Metasurface-aided {MIMO} Transceiver
  Design,'' \emph{IEEE Wireless Commun.}, vol.~31, no.~4, 2024.

\bibitem{AnJSAC2024}
J.~An, C.~Yuen, Y.~L. Guan, M.~Di~Renzo, M.~Debbah, H.~V. Poor, and L.~Hanzo,
  ``Two-Dimensional Direction-of-Arrival Estimation using Stacked Intelligent
  Metasurfaces,'' \emph{IEEE J. Sel. Areas Commun.}, vol.~42, no.~10, 2024.

\bibitem{AnJSAC2023}
J.~An, C.~Xu, D.~W.~K. Ng, G.~C. Alexandropoulos, C.~Huang, C.~Yuen, and
  L.~Hanzo, ``Stacked Intelligent Metasurfaces for Efficient Holographic {MIMO}
  Communications in {6G},'' \emph{IEEE J. Sel. Areas Commun.}, vol.~41, no.~8, 2023.

{

\bibitem{BayraktarAsilimoar2024}
M.~Bayraktar, N.~González-Prelcic, H.~Chen and C.~J.~Zhang,
  ``Near-Field Full-Duplex Integrated Sensing and Communication with Dynamic Metasurface Antennas,'' in \emph{Proc. IEEE 58th Asilomar Conference on Signals, Systems, and Computers (Asilomar CSSC)}, Pacific Grove, CA, USA, 2024.

  \bibitem{MatemuTWC2025}
  A. E. Matemu and K. Lee, ``Spatial Modulation and Generalized Spatial Modulation for Dynamic Metasurface Antennas,'' \emph{IEEE Trans. Wireless Commun.}, vol.~24, no.~1, Jan. 2025.

  \bibitem{Dardari_ARXIV_2024}
D.~Dardari, ``Dynamic Scattering Arrays for Simultaneous Electromagnetic Processing and Radiation in Holographic MIMO Systems,'' \emph{arXiv preprint:2405.16174}, 2024.

\bibitem{ShiTWC2025}
E.~Shi, J.~Zhang, Y.~Zhu, J.~An, C.~Yuen and B.~Ai, ``Uplink Performance of Stacked Intelligent Metasurface-Enhanced Cell-Free Massive MIMO Systems,'' \emph{IEEE Trans. Wireless Commun.}, 2025.

  \bibitem{StutzOJCOMS2025}
  A.~Stutz-Tirri, G.~Schwan and C.~Studer, ``Efficient and Physically Consistent Modeling of Reconfigurable Electromagnetic Structures,''
  \emph{IEEE Open J. Commun. Soc.}, vol.~6, 2025.

  \bibitem{AnTWC2025}
  J.~An, M.~D.~Renzo, M.~Debbah, H.~V.~Poor and C.~Yuen, ``Stacked Intelligent Metasurfaces for Multiuser Downlink Beamforming in the Wave Domain,'' \emph{IEEE Trans. Wireless Commun.}, 2025.

}

\bibitem{NeriniTWC2024}
M.~Nerini, S.~Shen, H.~Li, and B.~Clerckx, ``Beyond Diagonal Reconfigurable
  Intelligent Surfaces utilizing Graph Theory: Modeling, Architecture Design,
  and Optimization,'' \emph{IEEE Trans. Wireless Commun.}, vol.~23, no.~8, 2024.

\bibitem{WMA2024}
D.~Wijekoon, A.~Mezghani, G.~C. Alexandropoulos, and E.~Hossain,
  ``Physically-Consistent Modeling and Optimization of non-local {RIS}-Assisted
  Multi-User {MIMO} Communication Systems,'' \emph{arXiv preprint:2406.05617},
  2024.

\bibitem{Hadani_WCNC_2017}
R.~Hadani, S.~Rakib, M.~Tsatsanis, A.~Monk, A.~J. Goldsmith, A.~F. Molisch, and
  R.~Calderbank, ``Orthogonal Time Frequency Space Modulation,'' in \emph{Proc.
  IEEE Wireless Communications and Networking Conference (WCNC)}, San Francisco, USA, 2017.

\bibitem{Raviteja_TWC_2018}
P.~Raviteja, K.~T. Phan, Y.~Hong, and E.~Viterbo, ``Interference {C}ancellation
  and {I}terative {D}etection for {O}rthogonal {T}ime {F}requency {S}pace
  {M}odulation,'' \emph{IEEE Trans. Wireless Commun.}, vol.~17, no.~10, 2018.

\bibitem{Zhu_Arxiv23}
J.~Zhu, Y.~Tang, X.~Wei, H.~Yin, J.~Du, Z.~Wang, and Y.~Liu, ``A Low-Complexity
  Radar System based on Affine Frequency Division Multiplexing Modulation,''
  \emph{arXiv preprint arXiv:2312.11125}, 2023.

\bibitem{Liu_Arxiv24}
G.~Liu, T.~Mao, R.~Liu, and Z.~Xiao, ``Pre-Chirp-Domain Index Modulation for
  Affine Frequency Division Multiplexing,'' \emph{arXiv preprint
  arXiv:2402.15185}, 2024.

\bibitem{TseTIT2004}
D.~Tse, P.~Viswanath, and L.~Zheng, ``Diversity-Multiplexing Tradeoff in
  Multiple-Access Channels,'' \emph{IEEE Trans. Inf. Theory},
  vol.~50, no.~9, 2004.

\bibitem{MolischCOMMMAG2017}
A.~F. Molisch, V.~V. Ratnam, S.~Han, Z.~Li, S.~L.~H. Nguyen, L.~Li, and
  K.~Haneda, ``Hybrid Beamforming for Massive MIMO: A Survey,'' \emph{IEEE
  Commun. Mag.}, vol.~55, no.~9, 2017.

\bibitem{SandovalACCESS2023}
I.~A.~M. Sandoval, K.~Ando, O.~Taghizadeh, and G.~T.~F. De~Abreu, ``Sum-Rate
  Maximization and Leakage Minimization for Multi-User Cell-Free Massive MIMO
  Systems,'' \emph{IEEE Access}, vol.~11, 2023.

\bibitem{AnICC2024}
J.~An, C.~Yuen, Y.~L. Guan, M.~Di~Renzo, M.~Debbah, H.~V. Poor, and L.~Hanzo,
  ``Stacked Intelligent Metasurface performs a 2D DFT in the Wave Domain for
  DOA Estimation,'' in \emph{Proc. IEEE International Conference on Communications (ICC)}, Denver, CO, USA, 2024.

{

\bibitem{IimoriTWC2022}
H. Iimori, T. Takahashi, K. Ishibashi, G. T. F. de Abreu, D. González G. and O. Gonsa, ``oint Activity and Channel Estimation for Extra-Large MIMO Systems,'' \emph{IEEE Trans. Wireless Commun.}, vol.~21, no.~9, 2022.

\bibitem{ItoOJCOMS2024}
K. Ito, T. Takahashi, K. Furuta, S. Ibi and G. Thadeu Freitas de Abreu, ``Joint Channel and Data Estimation via Parametric Bilinear Inference for OTFS Demodulation,''
  \emph{IEEE Open J. Commun. Soc.}, vol.~5, 2024.

\bibitem{TakahashiTWC2024}
T. Takahashi, H. Iimori, K. Ishibashi, S. Ibi and G. T. F. de Abreu, ``Bayesian Bilinear Inference for Joint Channel Tracking and Data Detection in Millimeter-Wave MIMO Systems,'' \emph{IEEE Trans. Wireless Commun.}, vol.~23, no.~9, 2024.
}


\bibitem{TakahashiTCOM2019}
T.~Takahashi, S.~Ibi, and S.~Sampei, ``Design of Adaptively Scaled Belief in
  Multi-Dimensional Signal Detection for Higher-Order Modulation,'' \emph{IEEE
  Trans. Commun.}, vol.~67, no.~3, 2019.

\bibitem{Su_TSP_2015}
Q.~Su and Y.-C. Wu, ``On {C}onvergence {C}onditions of {G}aussian {B}elief
  {P}ropagation,'' \emph{IEEE Trans. Signal Process.}, vol.~63,
  no.~5, 2015.

{
\bibitem{Ranasinghe_BISTATIC_SIM_ARXIV_2025}
K.~R.~R. Ranasinghe, I.~A.~M. Sandoval, G.~T.~F. de~Abreu and G.~C. Alexandropoulos, ``Parametrized Stacked Intelligent Metasurfaces for Bistatic Integrated Sensing and Communications,'' \emph{arXiv preprint
  arXiv:2504.20661}, 2025.
}

\end{thebibliography}

\end{document}